\documentclass[11pt]{article}
\usepackage[a4paper, margin=27mm]{geometry}
\usepackage{newtxtext, newtxmath}
\usepackage{microtype}
\usepackage{amsmath}  
\usepackage{bm}
\usepackage{mathtools}
\usepackage{array}

\usepackage{graphicx} 

\begin{document}

\title{\vspace{-0.6cm}\textbf{Multivariate binary probability distribution in the Grassmann formalism}}

\author{Takashi Arai\thanks{\texttt{takashi-arai@sci.kj.yamagata-u.ac.jp}} \vspace{0.2cm} \\ 
Faculty of Science, Yamagata University, Yamagata 990-8560, Japan}


\date{\vspace{-0.5cm}}

\maketitle

\begin{abstract}
We propose a probability distribution for multivariate binary random variables.
The probability distribution is expressed as principal minors of the parameter matrix, which is a matrix analogous to the inverse covariance matrix in the multivariate Gaussian distribution.
In our model, the partition function, central moments, and the marginal and conditional distributions are expressed analytically.
That is, summation over all possible states is not necessary for obtaining the partition function and various expected values, which is a problem with the conventional multivariate Bernoulli distribution.
The proposed model has many similarities to the multivariate Gaussian distribution.
For example, the marginal and conditional distributions are expressed in terms of the parameter matrix and its inverse matrix, respectively.
That is, the inverse matrix represents a sort of partial correlation.
The proposed distribution can be derived using Grassmann numbers, anticommuting numbers.
Analytical expressions for the marginal and conditional distributions are also useful in generating random numbers for multivariate binary variables.
Hence, we investigated sampling distributions of parameter estimates using synthetic datasets.
The computational complexity of maximum likelihood estimation from observed data is proportional to the number of unique observed states, not to the number of all possible states as is required in the case of the conventional multivariate Bernoulli distribution.
We empirically observed that the sampling distributions of the maximum likelihood estimates appear to be consistent and asymptotically normal.
\end{abstract}

\section{Introduction}
The multivariate binary probability distribution is a model for multivariate binary random variables.
A well-known distribution for the multivariate binary variables is the multivariate Bernoulli distribution~\cite{Dai2013}, which is essentially the same as the Ising model in statistical physics~\cite{Ising1925}.
In the terminology of the graphical model, the multivariate Bernoulli distribution is a kind of Markov random field and is also called the Boltzmann machine in the field of machine learning research~\cite{Hinton1985}.
This model is used in many applications such as modeling the behavior of magnets in statistical physics, building statistical models in computer vision~\cite{Geman1984} and social network analysis.
Recent applications of this model include the study in detecting statistical dependence in the voting pattern from senate voting records data~\cite{Banerjee2008} and the study of cooperative mutations in the Human Immunodeficiency Virus (HIV)~\cite{Xue2012}.

The multivariate Bernoulli distribution encodes a binary variable as a dummy variable that takes discrete values in $\{0, 1\}$ or $\{-1, 1\}$.
However, the discrete nature of the dummy variables prevents us from analytical calculations.
For example, the marginal distribution is no longer in the same form as the original joint distribution.
Furthermore, a problem also arises from the viewpoint of computational complexity.
In the multivariate Bernoulli distribution, we have to sum over all possible states to calculate the partition function and various expected values; however, in a binary system, the number of possible states exponentially increases as the number of variables increases.
In other words, the computation of the partition function and expected values is $\mathrm{NP}$-hard, which causes difficulties with parameter estimation.
In fact, maximum likelihood estimation of model parameters by using a gradient-based method requires the calculation of various expected values, then, the application of such a usual estimation procedure becomes difficult when the number of variables is large.
In such a case, one way of dealing with parameter estimation is to approximate the expected values by Gibbs sampling, a Markov chain Monte Carlo simulation, but this method is computationally demanding and time-consuming.
Another way is to approximate the likelihood function to a more tractable functional form.
That is the variational inference~\cite{Wainwright2008}, the pseudolikelihood and the composition likelihood methods~\cite{Hofling2009, Ravikumar2010, Xue2012}, where methods for estimating the sparse structure of a graph are proposed through the use of $L_1$ and nonconcave regularizations.
Despite these efforts, the multivariate Bernoulli distribution has not been successful in practical application compared to the multivariate Gaussian distribution, whose partition function can be analytically computed and is successfully used in various fields such as natural language processing~\cite{Schutze1999}, image analysis~\cite{Woods1978, Hassner1980, Geman1984}, and spatial statistics~\cite{Ripley1981}.

In this paper, we propose a probability distribution that models multivariate binary variables.
To formulate the binary probability distribution, we use Grassmann numbers, anticommuting numbers.
Our model is based on the assumption that the summation over dummy variables can be replaced by the integration of Grassmann numbers.
The resulting model resolves the problem in the conventional multivariate Bernoulli distribution that summation over states cannot be calculated analytically.
The joint probability distribution is expressed as principal minors of the parameter matrix, which is a matrix analogous to the inverse covariance matrix in the multivariate Gaussian distribution.

This paper is organized as follows.
In Sec.~\ref{sec: proposed_method}, we summarize the proposed probability distribution. 
We also numerically verify that the distribution works in practice by demonstrating the sampling distributions of statistics.
In Sec.~\ref{sec: grassmann}, we derive the distribution using Grassmann numbers.
Readers who are interested in the application of the model rather than the theoretical background can safely skip this section.
In Sec.~\ref{sec: parameter_estimation}, we discuss a parameter estimation procedure from observed datasets and investigate the sampling distribution of maximum likelihood estimates using synthetic datasets of correlated binary variables.
Section~\ref{sec: conclusion} is devoted to conclusions.

\section{Proposed probability distribution ~\label{sec: proposed_method}}

\subsection{Statement of the result \label{sec: result} }
We consider a probability distribution for $p$-dimensional binary variables, each of which is coded by the dummy variables $x_i$ taking discrete values in $\{0, 1\}$.
The proposed distribution is expressed in terms of the $p \times p$ matrix of model parameters $\Sigma = \Lambda^{-1}$, which is analogous to the covariance matrix of the multivariate Gaussian distribution but not necessarily symmetric.
To discuss the joint distribution, we here define index labels for the variables.
We write the set of all indices of the $p$-dimensional binary variables as $P \equiv \{ 1, 2, \dots, p\}$.
Then, we write the index label for the variables observed as $x_i=1$ as $A \subseteq P$ and denote these variables as $x_A$.
In the same way, we write the index label for the variables observed as $x_i=0$ as $B \subseteq P$ and denote these variables as $x_B$.
Then, without loss of generality, the matrix of model parameters is represented by a partitioned matrix as follows:
\begin{align}
\Sigma = \begin{bmatrix} \Sigma_{AA} & \Sigma_{AB} \\ \Sigma_{BA} & \Sigma_{BB} \end{bmatrix}  
= \Lambda^{-1}
= \begin{bmatrix} \Lambda_{AA} & \Lambda_{AB} \\ \Lambda_{BA} & \Lambda_{BB} \end{bmatrix}^{-1} .
\end{align}
In this paper, we denote the proposed probability distribution by $\mathcal{G}$, named after ``Grassmann.''
The joint distribution is given by the principal minor of the parameter matrix as follows:
\begin{align}
p(x_A=1, x_B=0) =& \; \mathcal{G}\bigl(x_A=1, x_B=0 | \Sigma \bigr), \notag \\
\equiv & \; \frac{1}{\det \Lambda} \det( \Lambda_{BB} - I ) , \notag \\
=& \;  \det \begin{bmatrix} \Sigma_{AA} & -\Sigma_{AB} \\ \Sigma_{BA} & I - \Sigma_{BB} \end{bmatrix},  \label{eq: joint_distribution_p}
\end{align}
where $I$ denotes the identity matrix.
By using the dummy variables explicitly, the joint distribution can also be expressed as
\begin{align}
\mathcal{G}(x | \Sigma)
 = \det \begin{bmatrix} 
\Sigma_{11}^{x_1} ( 1- \Sigma_{11})^{1 - x_1} & \Sigma_{12} (-1)^{1 - x_2} & \Sigma_{13} (-1)^{1 - x_3} & \cdots \\
\Sigma_{21} (-1)^{1 - x_1} & \Sigma_{22}^{x_2} (1 - \Sigma_{22})^{1 - x_2} & \Sigma_{23} (-1)^{1 - x_3} & \cdots \\
\Sigma_{31} (-1)^{1 - x_1} & \Sigma_{32} (-1)^{1 - x_2} & \Sigma_{33}^{x_3} (1 - \Sigma_{33})^{1-x_3} & \cdots \\
\vdots & \vdots & \vdots & \ddots
\end{bmatrix}. \label{eq: joint_dummy}
\end{align}

For marginal distribution, we define the index labels of the marginalized and remaining variables as $M$ and $R$, the size of each index is $m$ and $p-m$, respectively.
The model parameters are represented by a partitioned matrix as follows:
\begin{align}
\Sigma = \begin{bmatrix} \Sigma_{RR} & \Sigma_{RM} \\ \Sigma_{MR} & \Sigma_{MM} \end{bmatrix}
= \Lambda^{-1} = \begin{bmatrix} \Lambda_{RR} & \Lambda_{RM} \\ \Lambda_{MR} & \Lambda_{MM} \end{bmatrix}^{-1}.
\end{align}
Then, the marginal distribution is expressed in terms of the principal submatrix of $\Sigma$, or the Schur complement of $\Lambda$ (see Appendix~\ref{sec: matrix_identity}):
\begin{align}
p(x_R) =&  \sum_{x_M \in \{0, 1\}^m} p(x_R, x_M), \notag \\
=& \; \mathcal{G}\bigl( x_R | \Sigma_{RR} \bigr), \\
\Sigma_{RR}  =& \bigl[ \Lambda_{RR} - \Lambda_{RM} \Lambda_{MM}^{-1} \Lambda_{MR} \bigr]^{-1}. \label{eq: marginal_p}
\end{align}

The conditional distribution is expressed as the Schur complement of $\Sigma$.
As in the case of the joint distribution, we write index labels for the variables observed as $x_i=1$ and $x_i=0$ as $A$ and $B$, and write these variables as $x_A$ and $x_B$, respectively.
We write the union of $A$ and $B$ as $C$, i.e., $x_C = (x_A, x_B)$.
Then, the remaining indices after conditioning are represented by the set difference by these indices $R = P \setminus (A \cup B) \equiv P \setminus C$.
Without loss of generality, the matrix of model parameters is represented by a partitioned matrix as follows:
\begin{align}
\Sigma = \begin{bmatrix} \Sigma_{RR} & \Sigma_{RC} \\ \Sigma_{CR} & \Sigma_{CC} \end{bmatrix}
= \begin{bmatrix} \Sigma_{RR} & \Sigma_{RA} & \Sigma_{RB} \\
\Sigma_{AR} & \Sigma_{AA} & \Sigma_{AB} \\
\Sigma_{BR} & \Sigma_{BA} & \Sigma_{BB} \end{bmatrix} =
 \Lambda^{-1} =& \begin{bmatrix} \Lambda_{RR} & \Lambda_{RC} \\ \Lambda_{CR} & \Lambda_{CC} \end{bmatrix}^{-1}
= \begin{bmatrix} \Lambda_{RR} & \Lambda_{RA} & \Lambda_{RB} \\
\Lambda_{AR} & \Lambda_{AA} & \Lambda_{AB} \\
\Lambda_{BR} & \Lambda_{BA} & \Lambda_{BB} \end{bmatrix}^{-1}.
\end{align}
Then, the conditional distribution is given by
\begin{align}
p(x_R | x_C) \equiv &\; p(x_R | x_A=1, x_B=0), \notag \\
=& \; \mathcal{G}\bigl( x_R | \Sigma_{R|x_C} \bigr), \\
\Sigma_{R|x_C} \equiv & \;  \Sigma_{RR} - \Sigma_{RC} \bigl[ \Sigma_{CC} - \mathrm{diag}(1 - x_C) \bigr]^{-1} \Sigma_{CR}, \notag  \\
= & \; \bigl[ \Lambda_{RR} - \Lambda_{RB} (  \Lambda_{BB} - I  )^{-1} \Lambda_{BR} \bigr]^{-1}, \label{eq: conditional_p}
\end{align}
where $\mathrm{diag}(1 - x_C)$ is the diagonal matrix with the diagonal elements given by $1 - x_C$:
\begin{align}
 \mathrm{diag}(1 - x_C) \equiv \; \delta_{ij} \, (1 - x_i), \hspace{0.5cm} (i, j \in C),  \label{eq: kronecker}
\end{align}
and $\delta_{ij}$ is the Kronecker delta.

The central moment for the variables with the index label $R  \subseteq P$ is given by
\begin{align}
m_R = & \, \mathrm{E}\Bigl[ (x_R - \mu_R) \Bigr], \notag \\
\equiv & \, \mathrm{E} \biggl[ \; \prod_{i \in R} (x_i - \mu_i ) \biggr], \notag \\
= & \, \det \Bigl[ \Sigma_{RR} - \mathrm{diag}(\mu_R) \Bigr], \label{eq: central_moment}
\end{align}
where $\mu_i = \mathrm{E}[x_i]$ is the mean parameter and $\mathrm{diag}(\mu_R)$ is a diagonal matrix defined by Eq.~(\ref{eq: kronecker}).

\subsection{Statistical properties and interpretation \label{sec: property} }
The diagonal elements of $\Sigma$ represent the mean of the marginal distribution of the dummy variables and must take a value in the range $[0, 1]$, and the product of the off-diagonal elements $- \Sigma_{ij} \Sigma_{ji}$ represent the covariance of the variables:
\begin{align}
\mathrm{E}[x_i] =& \Sigma_{ii}, \\
\mathrm{Cov}[x_i \, x_j] =& - \Sigma_{ij} \Sigma_{ji}. \label{eq: covariance}
\end{align}
Since uncorrelatedness between binary variables is equivalent to statistical independence, statistical independence is represented by the product of off-diagonal elements:
\begin{align}
\Sigma_{ij} \Sigma_{ji} = 0, \; \; \Leftrightarrow \; \; p(x_i, x_j) = p(x_i) p(x_j).
\end{align}

The diagonal element of $\Lambda$ represents the reciprocal of the mean of the dummy variable conditioned on all the other variables observed as $x_A = 1$:
\begin{align}
\mathrm{E}[x_i | x_A=1] = \Lambda_{ii}^{-1}, \hspace{0.5cm} (A = P \setminus i).
\end{align}
The product of the off-diagonal elements $\Lambda_{ij} \Lambda_{ji}$ is proportional to the partial covariance between $x_i$ and $x_j$ conditioned on all the other variables observed as $x_A = 1$:
\begin{align}
\mathrm{Cov}[x_i \, x_j | x_A =1]
=& \; \frac{ - \Lambda_{ij} \Lambda_{ji} }{ (\Lambda_{ii} \Lambda_{jj} - \Lambda_{ij}\Lambda_{ji} )^2}, \hspace{0.5cm} \bigl( A = P \setminus \{i, j\} \bigr). 
\end{align}
Therefore, the statistical independence between variables $x_i$ and $x_j$ conditioned on all the other variables observed as $x_A = 1$ is represented by the product of the off-diagonal elements:
\begin{align}
\Lambda_{ij} \Lambda_{ji} = 0, \; \; \Leftrightarrow \; \; p(x_i, x_j | x_A=1) = p(x_i | x_A = 1) \, p(x_j | x_A=1). \label{eq: conditional_independence}
\end{align}

In order for our model to make sense as a probability distribution, all principal minors of the matrix $\Lambda - I$ must be greater than or equal to zero.
In the terminology of linear algebra, the matrix $\Lambda - I$ must be a $P_0$ matrix.
Normalization of the probability distribution is satisfied by definition.
The normalization constant, that is, the partition function, is given by the matrix determinant, $\det \Lambda$.

Our formalism does not depend on how to encode binary variables, i.e., the dummy variable has flip symmetry.
The dummy variable $x_i$ with the mean and covariance given by $\mu_i$ and $-\Sigma_{ij} \Sigma_{ji} \; \;  ( j \ne i)$, is equivalent to the dummy variable $\tilde{x}_i \equiv 1 - x_i$ with the mean and covariance given by $1 - \mu_i$ and $-\Sigma_{ij} (- \Sigma_{ji} )$.
This flip symmetry can also be read from the expression of the joint distribution, Eq.~(\ref{eq: joint_dummy}), which uses the dummy variable explicitly.
The parameter matrix with the dummy coding flipped is expressed as (see Appendix~\ref{sec: matrix_identity})
\begin{align}
\tilde{\Sigma}^{-1} \equiv & \begin{bmatrix} \Sigma_{AA} & -\Sigma_{AB} \\ \Sigma_{BA} & I - \Sigma_{BB} \end{bmatrix}^{-1} , \notag \\
=& \begin{bmatrix}
\Lambda_{AA} - \Lambda_{AB} (\Lambda_{BB} - I)^{-1} \Lambda_{BA} & - \Lambda_{AB} (\Lambda_{BB} - I)^{-1} \\
 (\Lambda_{BB} - I)^{-1} \Lambda_{BA} & \Lambda_{BB} (\Lambda_{BB} - I)^{-1} 
\end{bmatrix} \equiv  \tilde{\Lambda}. \label{eq: flip_symmetry}
\end{align}
Hence, the interpretation of conditional statistical independence, such as Eq.~(\ref{eq: conditional_independence}), is applicable to general conditioning other than the conditioning of the variables as $x_A=1$, by redefining the parameter matrix as $\tilde{\Lambda}$.

The parameter matrices $\Sigma$ and $\Lambda$ have redundant degrees of freedom.
That is, there exist different parameters that generate exactly the same joint probability.
In fact, the joint distribution is invariant under multiplying an $i$th row of the matrix $\Sigma$ by a constant $c_i$ at the same time multiplying the $i$th column with the same index $i$ by the constant $1/c_i$.
Furthermore, the joint distribution is invariant under the matrix transposition.
These degrees of freedom are expressed in the form of a matrix as
\begin{align}
\Sigma' =& D^{-1} \Sigma D, \hspace{0.2cm} \Leftrightarrow \hspace{0.2cm}  \Lambda' = D^{-1} \Lambda D, \label{eq: dilatation} \\
\Sigma' =& \Sigma^T, \hspace{0.2cm} \Leftrightarrow \hspace{0.2cm} \Lambda'  = \Lambda^T , \label{eq: transposition}
\end{align}
where $D$ is an arbitrary invertible diagonal matrix.

The proposed distribution has many similarities to the multivariate Gaussian distribution.
In the multivariate Gaussian distribution, the covariant structure of the joint distribution is described by the covariance matrix.
Furthermore, the covariance structure of the marginal and conditional distributions are described by the submatrix and the Schur complement of the covariance matrix, respectively.
The presence or absence of correlation in the marginal and conditional distribution is equivalent to unconditional and conditional statistical independence, respectively.
These properties also hold in our model.
However, our model differs from the multivariate Gaussian distribution in the following ways.
First, the parameter matrix $\Sigma=\Lambda^{-1}$ is generally not symmetric.
The covariance between dummy variables for the marginal distribution is given by the product of the off-diagonal elements, Eq.~(\ref{eq: covariance}), that is, positive covariances are achieved by setting $\Sigma_{ij}$ and $\Sigma_{ji}$ with different signs.
The covariance of a conditional distribution, Eq.~(\ref{eq: conditional_p}), depends on the observed values of the conditioning variables.
However, it is plausible to think that this is because the mean can be shifted by conditioning and, in the binary variables, the mean and variance are linked to each other.
Maximum likelihood estimation of the model parameters from observed data has to resort to numerical calculations, which will be discussed in Sec.~\ref{sec: parameter_estimation}, while in the multivariate Gaussian distribution, the maximum likelihood estimator of the covariance matrix is given by the sample covariance.

\subsection{Sampling distribution of statistics}
In this section, we numerically demonstrate that our model works in practice by generating random numbers and investigating sampling distributions of statistics.
Since the analytical expressions for the marginal and conditional distributions are obtained in our formalism, we can easily generate correlated random numbers for multivariate binary variables by repeating Bernoulli trials.
In fact, since the joint distribution can be factorized as $p(x_1, x_2, \dots, x_p) = p(x_1) p(x_2 | x_1) \cdots p(x_p | x_1, x_2, \dots, x_{p-1})$, we can generate a random number by repeating Bernoulli trials $p$ times from $p(x_1)$ to $p(x_p | x_1, x_2, \dots, x_{p-1})$ depending on the previous observations.
We investigate the sampling distributions of various statistics given the model parameters.
The parameters are chosen based on the maximum entropy principle as discussed in Sec.~\ref{sec: parameter_estimation}, subject to having a specified mean and covariance.
The parameter matrix is given by
\begin{align}
\Sigma = \begin{bmatrix}
0.85 & -0.34 & -0.07 & 0.16 & -0.06 \\
-0.11 & 0.46 & 0.06 & -0.09 & -0.05 \\
-0.16 & -0.42 & 0.74 & 0.66 & -0.28 \\
0.01 & -0.08 & -0.13 & 0.70 & -0.30 \\
0.02 & 0.15 & -0.04 & 0.23 & 0.80
\end{bmatrix} \simeq
\begin{bmatrix}
1.30 & 0.91 & 0.01 & -0.22 & 0.08 \\
0.26 & 2.25 & -0.09 & 0.23 & 0.22 \\
0.37 & 1.09 & 1.12 & -1.04 & 0.10 \\
0.05 & 0.25 & 0.21 & 1.09 & 0.50 \\
-0.08 & -0.47 & 0.02 & -0.40 & 1.07
\end{bmatrix}^{-1} = \Lambda^{-1}, \label{eq: maximum_solution}
\end{align}
where the value of each element is rounded for presentation.

\begin{figure}[htbp]
\centering
\includegraphics[width=14cm]{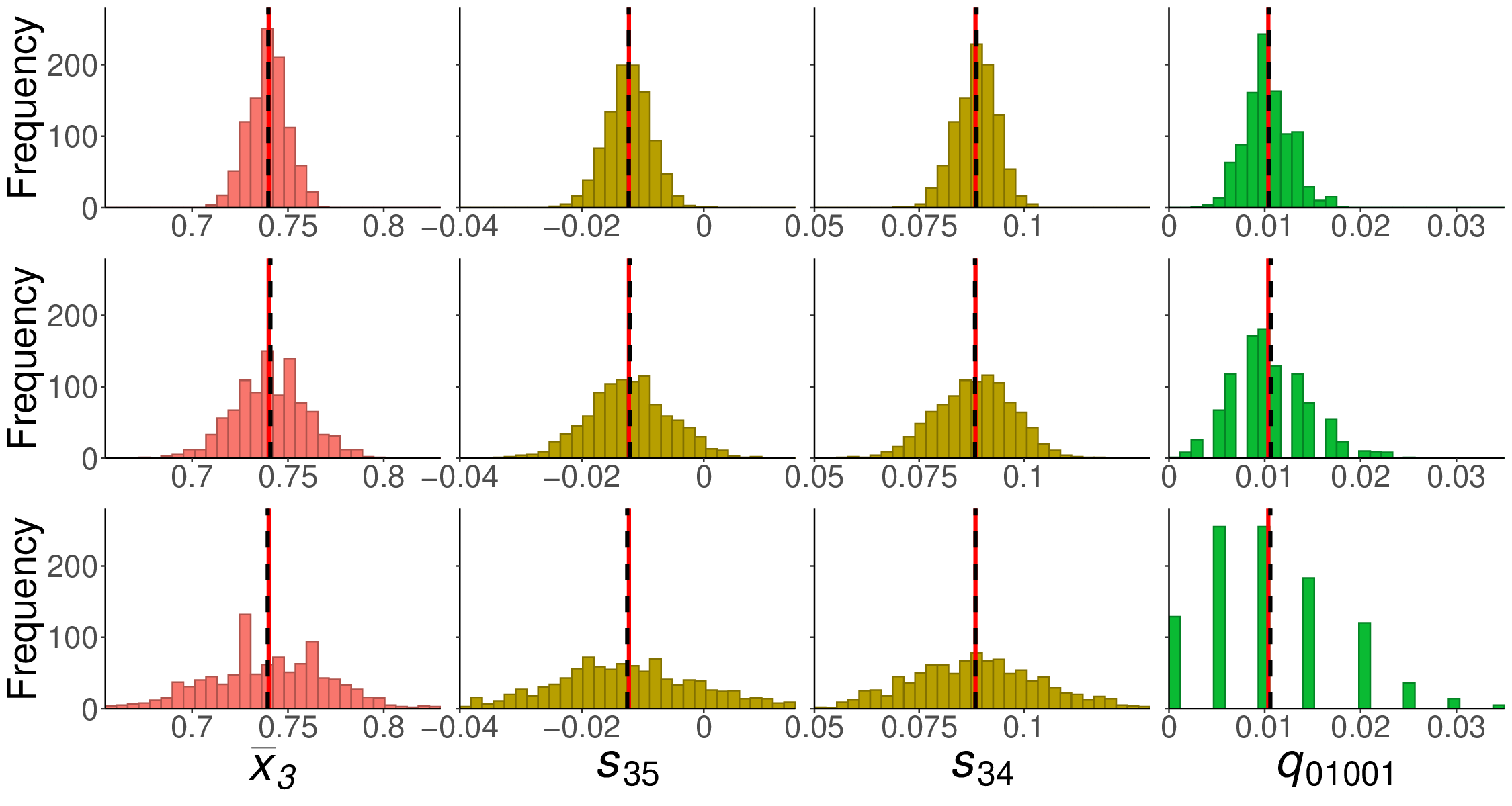}
\caption{Examples of the sampling distributions of the statistics for the sample mean $\bar{x}_3$, unbiased sample covariances $s_{35}$, $s_{34}$, and the empirical joint probability $q(x_P=(0,1,0,0,1))$.
The first, second and third rows correspond to the results of the sample sizes $N=2000$, $N=600$, and $N=200$, respectively.
The trial size, i.e., the number of times the statistics are computed, is $M=1000$.
The red solid lines denote the true values and the black dashed lines denote the mean values of the sampling distributions.
} 
\label{fig: sampling_distribution_statistics}
\end{figure}

Figure~\ref{fig: sampling_distribution_statistics} shows the sampling distributions of the statistics for the sample mean and unbiased sample covariance and the empirical joint distribution from the synthetic datasets for different sample sizes $N$.
The unbiased sample covariance is defined as
\begin{align}
s_{ij} \equiv & \; \frac{1}{N-1} \sum_{n=1}^N (x_{ni} - \bar{x}_i ) (x_{nj} - \bar{x}_j), \hspace{0.5cm} ( i \ne j), \\
\bar{x}_i \equiv & \; \frac{1}{N} \sum_{n=1}^N x_{ni}.
\end{align}
We observe that the sampling distributions of the statistics are consistent with the theoretical predictions, e.g., the mean and variance of the statistics obey
\begin{align}
\mathrm{E}[ \bar{x}_i] =& \, \mu_i, \\
\mathrm{Var}[ \bar{x}_i] =& \, \frac{1}{N} \, \mu_i \, (1 - \mu_i), 
\end{align}
and
\begin{align}
\hspace{3.5cm} \mathrm{E}[ s_{ij} ] =&  - \Sigma_{ij} \Sigma_{ji} , \notag \\
\equiv & \; \sigma_{ij} , \\
\mathrm{Var}[s_{ij} ] =& \frac{1}{N} m_{ iijj } - \frac{ (N - 2)}{ N (N-1)} \bigl( m_{ ij } \bigr)^2 + \frac{1}{N(N-1)} m_{ ii } \, m_{ jj },
\end{align}
where $m_{ ii }$ and $m_{ iijj }$ are the second- and fourth-order central moments, respectively:
\begin{align}
m_{ii} =& \mu_i (1 - \mu_i), \\
m_{ iijj } \equiv & \, \mathrm{E} \Bigl[ (x_i - \mu_i)^2 (x_j - \mu_j)^2 \Bigr], \; \; \; (i \ne j), \notag \\
= & \, \mathrm{E} \Biggl[ \Bigl[ (1 - 2 \mu_i) x_i + \mu_i^2 \Bigr] \Bigl[ (1 - 2 \mu_j) x_j + \mu_j^2 \Bigr] \Biggr] , \notag \\
=&  (1 - 2\mu_i)(1-2 \mu_j) (- \Sigma_{ij} \Sigma_{ji} ) + \mu_i (1 -\mu_i) \mu_j (1 - \mu_j).
\end{align}
Although the sampling distribution can be skewed when the sample size is small, it becomes asymptotically normal as the sample size increases, which is consistent with the central limit theorem.

\section{Derivation using Grassmann numbers \label{sec: grassmann} }
In this section, we derive the proposed probability distribution using Grassmann numbers.
The properties of Grassmann numbers can be consulted in the Appendix~\ref{sec: grassmann_appendix}.
Readers who are interested in the application of the model rather than the theoretical background can safely skip this section.

The multivariate Bernoulli distribution is a probability distribution for binary random variables, where $p$-dimensional binary variables are encoded by the vector of dummy variables $x_P$ taking $x_i \in \{0, 1\} , \;\;i \in P \equiv \{1,2,\dots, p\}$.
Usually, the joint distribution of the multivariate Bernoulli distribution is expressed as an exponential function of a polynomial up to the second order for the dummy variables~\cite{Wainwright2008, Dai2013},
\begin{align}
p(x_P) = \frac{1}{Z} \exp \biggl\{ \sum_{i=1}^p b_i \, x_i + \sum_{i, \,j=1}^p x_i \, w_{ij} \, x_j \biggr\}, \label{eq: Bernoulli_distribution}
\end{align}
where $b_i$ and $w_{ij}$ are called the bias and weight terms, and the exponent is called the energy function.
$Z$ is the partition function that ensures that the distribution sums to one.
In the conventional multivariate Bernoulli distribution, various quantities such as the partition function and expected values are computed by summation over all possible states.
For example, the expected value of the random variable $x_i$ is expressed as the following summation:
\begin{align}
\mathrm{E}[x_i] =& \sum_{x_P \in \{0, 1\}^p} x_i \, p(x_P), \notag \\
=& \sum_{x_1 \in \{0,1\}} \sum_{x_2 \in \{0,1\}} \cdots \sum_{x_p \in \{0,1\}}  x_i \, p(x_1, x_2, \dots, x_p).
\end{align}
However, for multivariate binary variables, the number of possible states increases exponentially as the dimension of the variable $p$ increases.
Then, performing the summation over states becomes difficult even numerically.
Furthermore, there exists a difficulty with the conventional multivariate Bernoulli distribution that the marginal distributions do not follow the multivariate Bernoulli distribution, though the conditional distributions still follow the multivariate Bernoulli distribution.
In fact, the marginal distribution $p(x_R)$ for indices $R = P \setminus M$, in which the variables with the indices $M \subseteq P$ are marginalized out,
\begin{align}
p(x_R) = \sum_{x_M} p(x_R, x_M),
\end{align}
is no longer in the same form as the original expression of Eq.~(\ref{eq: Bernoulli_distribution}).
Then, it is difficult to interpret the model parameters, which is in contrast to the multivariate Gaussian distribution where the covariance matrix and its inverse matrix can be interpreted as indirect and direct correlations.
We try to resolve these difficulties by introducing Grassmann numbers, anticommuting numbers.
We introduce a pair of Grassmann variables $\theta$, $\bar{\theta}$ corresponding to the dummy variables $x_P$.
We make an Ansatz that the summation over states can be replaced by the integration of Grassmann variables.
Then, we expect that the partition function and various expected values can be expressed analytically.

\subsection{Univariate binary probability distribution}
First, we explain our idea with the simplest example of the univariate binary probability distribution.
In the conventional Bernoulli distribution, the normalization condition of the probability distribution and expected values of the random variable are computed by summation over all possible states of the dummy variable $x \in \{ 0, 1\}$:
\begin{align}
\sum_x p(x) =& 1, \notag \\
\mathrm{E}[x] =& \sum_x x \, p(x) = p(x=1) .
\end{align}
On the other hand, our formalism introduces a pair of Grassmann variables $\theta$ and $\bar{\theta}$~\cite{Peskin1995}, anticommuting numbers, corresponding to the dummy variable.
These variables obey the following anticommutation relations (see Appendix~\ref{sec: grassmann_appendix}):
\begin{align}
\{ \theta, \bar{\theta} \} \equiv& \;  \theta \bar{\theta} + \bar{\theta} \theta = 0, \notag \\
\theta^2 =& \; \bar{\theta}^2 = 0.
\end{align}
Then, we assume that instead of the summation described above expected values can be obtained by integration of the Grassmann function defined by
\begin{align}
\frac{1}{Z(\Lambda)} \; e^{H(\bar{\theta}, \theta)} =& \;  \frac{1}{Z(\Lambda)} \; e^{\bar{\theta} \Lambda \theta },
\end{align}
where $\Lambda$ is a parameter of the model, and $Z(\Lambda)$ is the partition function, the normalization constant that ensures that the distribution sums to one.
We hereafter refer to the exponent of the Grassmann function as Hamiltonian $H(\bar{\theta}, \theta)$.
In the above equation, we have adopted the quadratic form in the Grassmann variables as a Hamiltonian.

The Ansatz that the summation over dummy variables can be replaced by the integration of Grassmann variables can be confirmed by direct calculation.
In fact, the partition function is calculated by integration of the Grassmann variables as follows:
\begin{align}
Z(\Lambda) =& \int d\theta d\bar{\theta} \, e^{\bar{\theta} \Lambda \theta}, \notag \\
= & \Lambda \equiv \frac{1}{\Sigma}.
\end{align}
The expected value of the dummy variable $x$, which corresponds to the probability of $p(x=1)$, is consistently calculated as the expected value of the product of the Grassmann variables $(\bar{\theta} \theta)$ as follows:
\begin{align}
\mathrm{E}[x] = \; p(x=1) 
= &  \frac{1}{Z(\Lambda)}  \int d\theta d\bar{\theta} \, (\bar{\theta} \theta) \,  e^{H(\bar{\theta}, \theta)}, \notag \\
=& \;  \frac{1}{\Lambda} \int d\theta d\bar{\theta} \,  (\bar{\theta} \theta) \, e^{\bar{\theta} \Lambda \theta}, \notag \\
=&\;  \frac{1}{\Lambda} \equiv \Sigma.
\end{align}
The probability $p(x=0)$ is calculated by the expected value of the Grassmann variables $(1-\bar{\theta}\theta)$,
\begin{align}
p(x=0) =&  \frac{1}{Z(\Lambda)} \int d\theta d\bar{\theta} \, (1- \bar{\theta}\theta) \, e^{H(\bar{\theta}, \theta)}, \notag \\
=& 1 -\Sigma,
\end{align}
which is an analogy from the summation $p(x=0) = \sum_x (1-x) \, p(x)$.
Thus, we see that the parameter $\Sigma$ can be interpreted as the mean parameter of the probability distribution.
Higher-order central moments can be derived consistently by the following prescription.
Since the higher-order terms of Grassmann variables vanish, we first summarize the polynomials for the dummy variable in the central moment using the identity, $x^k = x, \; k \in \{ 1, 2, \dots \}$:
\begin{align}
\mathrm{E}\bigl[ (x - \mu)^k \bigr] =\,  \mathrm{E}[x] \, \sum_{l=0}^{k-1} \binom{k}{l} \, (-\mu)^l + (-\mu)^k.
\end{align}
Then, the Grassmann integral for the above expression gives consistent results.
Therefore, our formalism successfully reproduces the univariate Bernoulli distribution.

\subsection{Bivariate binary probability distribution}
The same idea as the previous subsection is applicable to the bivariate binary probability distribution.
We introduce a pair of Grassmann vectors $\theta = (\theta_1, \theta_2)^T $, $\bar{\theta} = (\bar{\theta}_1, \bar{\theta}_2)^T$ corresponding to the dummy variables $x_1$, $x_2$.
Again, we make an Ansatz that the expected value by summation over states can be calculated by integration of the following exponential function of the Grassmann variables,
\begin{align}
\frac{1}{Z(\Lambda)} \; e^{H(\bar{\theta}, \theta)} = \frac{1}{Z(\Lambda)} \, e^{\theta^{\dagger} \Lambda \theta},
\end{align}
where $\theta^{\dagger}$ denotes the transpose of the Grassmann vector $\bar{\theta}$, $\theta^{\dagger} \equiv \bar{\theta}^T$, and $\Lambda = \Sigma^{-1}$ is a matrix of model parameters analogous to the precision and covariance matrices in the bivariate Gaussian distribution,
\begin{align}
\Lambda = \begin{bmatrix} \Lambda_{11} & \Lambda_{12} \\ \Lambda_{21} & \Lambda_{22} \end{bmatrix} 
= \Sigma^{-1} = \begin{bmatrix} \Sigma_{11} & \Sigma_{12} \\ \Sigma_{21} & \Sigma_{22} \end{bmatrix}^{-1}.
\end{align}
By performing the Grassmann integral, the partition function is represented by the determinant of the matrix $\Lambda$,
\begin{align}
Z(\Lambda) =& \int d\theta_1 d\bar{\theta}_1 d\theta_2 d\bar{\theta}_2 \, e^{\theta^{\dagger} \Lambda \theta} , \notag \\
=& \det \Lambda.
\end{align}

We first confirm that our Ansatz reproduces the joint distribution.
In the conventional bivariate Bernoulli distribution, the co-occurrence probability $p(x_1 = 1, x_2 = 1)$ can be rewritten as an expected value of the dummy variables,
\begin{align}
p(x_1=1, x_2=1) =& \sum_{x_1, x_2} x_1 x_2 \, p(x_1, x_2), \notag \\
=&  \; \mathrm{E}[x_1 \, x_2].
\end{align}
In our formalism, the above summation over states is expressed as the Grassmann integral.
In fact, the co-occurrence probability is calculated as
\begin{align}
p(x_1 = 1, x_2=1) =& \frac{1}{ \det \Lambda} \int d\theta_1 d\bar{\theta}_1 d\theta_2 d\bar{\theta}_2 \, (\bar{\theta}_1 \theta_1)
(\bar{\theta}_2 \theta_2) \, e^{\theta^{\dagger} \Lambda \theta} , \notag \\
= & \frac{1}{\det \Lambda} .
\end{align}
In the same way, the joint probabilities of the remaining states are calculated as
\begin{align}
p(x_1 = 1, x_2=0) 
= & \; \frac{\Lambda_{22} - 1}{\det \Lambda}, \\
p(x_1 = 0, x_2=1)
= & \;  \frac{\Lambda_{11} - 1 }{\det \Lambda}, \\
p(x_1 = 0, x_2=0) 
= & \;  \frac{\det (\Lambda - I)}{\det \Lambda},
\end{align}
where $I$ is the identity matrix.
The above expressions for the joint distribution can also be interpreted as all of the principal minors of the matrix $\Lambda - I$ divided by $\det \Lambda$.
By using the dummy variables explicitly, the joint probabilities are summarized in terms of $\Sigma$ as
\begin{align}
p(x_1, x_2) = \det \begin{bmatrix}
\Sigma_{11}^{x_1} (1 - \Sigma_{11} )^{1- x_1} & (-1)^{1-x_2} \Sigma_{12} \\
(-1)^{1-x_1} \Sigma_{21} & \Sigma_{22}^{x_2} (1 - \Sigma_{22})^{1 - x_2} \end{bmatrix}. \label{eq: joint}
\end{align}

Next, we turn to the marginal distribution.
In the conventional bivariate Bernoulli distribution, marginalization of the variable $x_2$ is taken by the summation of the dummy variable:
\begin{align}
p(x_1) = \sum_{x_2} p(x_1, x_2).
\end{align}
Again, the above marginalization can be performed by the integration of the Grassmann variables $\theta_2$ and $\bar{\theta_2}$, which is calculated by completing the square and shifting the integral variables as
\begin{align}
\frac{1}{Z(\Lambda)} \int d\theta_2 d\bar{\theta}_2 \, e^{H(\bar{\theta}, \theta)}
=  & \frac{1}{\det \Lambda} \int d\theta_2 d\bar{\theta}_2 \exp \Bigl\{\theta^{\dagger} \Lambda \theta \Bigr\},  \notag \\
=& \frac{1}{\det \Lambda} \int d\theta_2 d\bar{\theta}_2 \exp \Bigl\{ \bar{\theta}_1 ( \Lambda_{11} - \Lambda_{12} \Lambda_{22}^{-1} \Lambda_{21} )  \theta_1 \notag \\
& \hspace{3.6cm} + (\bar{\theta}_2 + \bar{\theta}_1 \Lambda_{12} \Lambda_{22}^{-1}) \Lambda_{22} (\theta_2 + \Lambda_{22}^{-1} \Lambda_{21} \theta_1) \Bigr\} , \notag \\
=& \frac{\Lambda_{22}}{\det \Lambda} \exp \Bigl\{ \bar{\theta}_1 (\Lambda_{11} - \Lambda_{12} \Lambda_{22}^{-1} \Lambda_{21} ) \theta_1 \Bigr\}, \notag \\
\equiv & \frac{1}{Z_{x_1}} \; e^{H( \bar{\theta}_1, \theta_1)}.
\end{align}
Here we shall call the resulting Hamiltonian $H(\bar{\theta}_1, \theta_1)$ the marginal Hamiltonian.
From the above expression, we can read that the marginal distribution still follows the same form as the original joint distribution and the parameter of the resulting distribution is the Schur complement of the matrix $\Lambda$.
In terms of $\Sigma$, the marginal distribution is simply expressed as
\begin{align}
\frac{1}{Z_{x_1}} \; e^{H(\bar{\theta}_1, \theta_1)} = \Sigma_{11} \, e^{ \bar{\theta}_1 \Sigma_{11}^{-1} \theta_1}.
\end{align}
Therefore, the diagonal elements of the matrix $\Sigma$ can be interpreted as the mean parameters of the marginal distributions.

Here we discuss the correlation, covariance, and statistical independence of the variables.
The covariance between $x_1$ and $x_2$ can be calculated by the Grassmann integral as
\begin{align}
\mathrm{Cov}[x_1 \, x_2] =& \; \mathrm{E}[(x_1 - \mu_1) (x_2 - \mu_2)] , \notag \\
=& \; \frac{1}{\det \Lambda} \int d\theta_1 d \bar{\theta}_1 d\theta_2 d \bar{\theta}_2 \, ( \bar{\theta}_1 \theta_1 - \mu_1 )( \bar{\theta}_2 \theta_2 - \mu_2 ) \, e^{\theta^{\dagger} \Lambda \theta} ,\notag \\
=& - \Sigma_{12} \Sigma_{21}, 
\end{align}
where $\mu_i \equiv \mathrm{E}[x_i] = \Sigma_{ii}$ is the mean parameter.
Therefore, the product of the off-diagonal elements can be interpreted as the covariance of the variables.
Here, we define the correlation of binary variables by the Pearson correlation coefficient expressed as
\begin{align}
\rho_{12} \equiv & \frac{\mathrm{Cov}[x_1 \, x_2] }{\sqrt{ \mathrm{Cov}[x_1 \, x_1] \mathrm{Cov}[x_2 \, x_2] }} , \notag \\
=& \frac{ - \Sigma_{12} \Sigma_{21} }{ \sqrt{\Sigma_{11} (1 - \Sigma_{11}) \Sigma_{22} (1- \Sigma_{22}) } }. \label{eq: pearson_correlation}
\end{align}
Then, we notice that the expression for the joint distribution, Eq.~(\ref{eq: joint}), can be transformed to
\begin{align}
p(x_1, x_2) = & p(x_1) p(x_2) - (-1)^{x_1 + x_2} \Sigma_{12} \Sigma_{21}, \notag \\
=& p(x_1) p(x_2) + (-1)^{x_1 + x_2}  \sqrt{\Sigma_{11} (1 - \Sigma_{11}) \Sigma_{22} (1- \Sigma_{22}) } \;  \rho_{12}.
\end{align}
The above equation confirms that the uncorrelatedness between variables $x_1$ and $x_2$ is equivalent to statistical independence.

Last, we discuss the conditional distribution.
In the conventional Bernoulli distribution, the conditioning on the observation $x_2=1$ is expressed as a summation over the dummy variable through the Bayes' theorem:
\begin{align}
p(x_1 | x_2=1) =& \frac{p(x_1, x_2=1)}{p(x_2=1)} ,\notag \\
=& \frac{\sum_{x_2} x_2 \, p(x_1, x_2)}{ p(x_2=1) }.
\end{align}
Again, the above summation is rewritten by the Grassmann integral.
The Hamiltonian corresponding to the conditional distribution, which we call the conditional Hamiltonian $H(\bar{\theta}_1, \theta_1 | x_2=1)$, is calculated as
\begin{align}
\frac{1}{Z_{x_1 | x_2=1}} \; e^{H(\bar{\theta}_1, \theta_1 | x_2 = 1)} \equiv & \; \frac{1}{p(x_2 = 1)}  \frac{1}{Z(\Lambda)} \int d\theta_2 d\bar{\theta}_2 \, (\bar{\theta}_2 \theta_2) \, e^{H(\bar{\theta}, \theta)} , \notag \\
=& \;  \frac{1}{p(x_2 = 1)} \frac{1}{\det \Lambda} \int d\theta_2 d\bar{\theta}_2 \, (\bar{\theta}_2 \theta_2) \, e^{ \theta^{\dagger} \Lambda \theta} , \notag \\
=& \;  \frac{1}{\Lambda_{11} } \; e^{ \bar{\theta}_1 \Lambda_{11} \theta_1} .
\end{align}
Therefore, the conditional distribution given $x_2 = 1$ still follows the same form as the original joint distribution and the model parameter is just the principal submatrix of $\Lambda$.
The above conditional distribution can also be expressed in terms of the Schur complement of $\Sigma$ with respect to $\Sigma_{22}$:
\begin{align}
\frac{1}{Z_{x_1 | x_2=1}} \; e^{H(\bar{\theta}_1, \theta_1 | x_2 = 1)} =& \, \Sigma_{1|2} \; e^{ \bar{\theta}_1 \Sigma_{1|2}^{-1} \theta_1}, \\
\Sigma_{1|2} =& \, \Sigma_{11} - \Sigma_{12} \Sigma_{22}^{-1} \Sigma_{21}.
\end{align}
In the same way, the conditional Hamiltonian by the observation $x_2=0$ is calculated as
\begin{align}
\frac{1}{Z_{x_1 | x_2=0}} \; e^{H(\bar{\theta}_1, \theta_1 | x_2=0)} \equiv & \; \frac{1}{p(x_2=0)} \frac{1}{\det \Lambda} \int d\theta_2 d\bar{\theta}_2 \, (1-\bar{\theta}_2 \theta_2) \,  e^{ \theta^{\dagger} \Lambda \theta}, \notag \\
=& \; \frac{1}{\det \Lambda - \Lambda_{11}} \int d\theta_2 d\bar{\theta}_2 \, e^{-\bar{\theta}_2 \theta_2} \,  e^{ \theta^{\dagger} \Lambda \theta}, \notag \\
=& \; \frac{1}{\Lambda_{11} - \Lambda_{12}(\Lambda_{22} - 1)^{-1} \Lambda_{21} } \exp \Bigl\{ \bar{\theta}_1 \bigl[ \Lambda_{11} - \Lambda_{12} (\Lambda_{22} - 1)^{-1} \Lambda_{21} \bigr] \theta_1 \Bigr\}.
\end{align}
Again, the conditional distribution given $x_2=0$ still follows the same form as the joint distribution.
From the above conditional distribution, we can read the symmetry of dummy coding in our formalism.
In fact, the mean of the variable $x_1$ is shifted by each conditioning as follows:
\begin{align}
p(x_1=1 | x_2=1) =& \; \Sigma_{11} - \Sigma_{12} \Sigma_{22}^{-1} \Sigma_{21}, \\
p(x_1=1 | x_2=0) =& \; \Sigma_{11} - \Sigma_{12} (\Sigma_{22} - 1)^{-1} \Sigma_{21}, \notag \\
=& \; \Sigma_{11} - (- \Sigma_{12} ) (1 - \Sigma_{22} )^{-1} \Sigma_{21}.
\end{align}
The conditional distribution given $x_2=1$ is expressed as the partial covariance matrix $\Sigma_{1|2}$ for observing the variable with the mean $\Sigma_{22}$ and covariance $-\Sigma_{12}\Sigma_{21}$.
On the other hand, the conditional distribution given $x_2=0$ is expressed as the partial covariance matrix for observing variables with the mean and the sign of the correlation are inverted as $1 - \Sigma_{22}$ and $- ( - \Sigma_{12}) \Sigma_{21}$.
In other words, observing the dummy variable $x_2$ as $x_2=0$ is equivalent to observing the dummy variable $\tilde{x}_2 \equiv 1 - x_2 $ with the dummy coding inverted as $\tilde{x}_2 = 1$.
The conditional distribution given $x_2=0$ is simply expressed as the Schur complement of the matrix $\tilde{\Sigma}$, 
\begin{align}
\tilde{\Sigma} \equiv \begin{bmatrix} \Sigma_{11} & - \Sigma_{12} \\ \Sigma_{21} & 1 - \Sigma_{22} \end{bmatrix},
\end{align}
with respect to $\tilde{\Sigma}_{22} = 1 - \Sigma_{22}$.

\subsection{$p$-dimensional binary probability distribution}
The procedure in the previous subsections can be extended to $p$-dimensional variables straightforwardly.
In this subsection, we enumerate the results.
First, we introduce a pair of $p$-dimensional Grassmann vectors $\theta=(\theta_1, \theta_2, \dots, \theta_p)^T $, $\theta^{\dagger} = \bar{\theta}^T = (\bar{\theta}_1, \bar{\theta}_2, \dots, \bar{\theta}_p) $ and $p \times p$ matrix of model parameters, $\Lambda = \Sigma^{-1}$.
Then, we introduce the following Hamiltonian:
\begin{align}
\frac{1}{Z(\Lambda)} \; e^{H(\bar{\theta}, \theta)} \equiv \frac{1}{\det \Lambda} \, e^{\theta^{\dagger} \Lambda \theta}.
\end{align}
The probability distribution is defined by the integral of the above Grassmann function.
We adopt the following sign convention of the Grassmann integral:
\begin{align}
\int d\theta_1 d\bar{\theta}_1 d\theta_2 d\bar{\theta}_2 \cdots d\theta_p d\bar{\theta}_p \, (\bar{\theta}_1 \theta_1 ) (\bar{\theta}_2 \theta_2 ) \cdots (\bar{\theta}_p \theta_p ) = 1.
\end{align}

We define the index labels for the variables observed as $x_i=1$ and $x_i=0$ as $A$ and $B$, respectively, as defined by Sec.~\ref{sec: result}.
Then, the joint distribution is given by
\begin{align}
p(x_A=1, x_B=0) =& \; \frac{1}{\det \Lambda} \int d\theta_P d\bar{\theta}_P
\, (\bar{\theta}_A \theta_A) (1 - \bar{\theta}_B \theta_B) \, e^{\theta^{\dagger} \Lambda \theta} , \notag \\
\equiv & \;
\frac{1}{\det \Lambda} \int \biggl[ \prod_{i=1}^p d\theta_i d\bar{\theta}_i \biggr] \biggl[ \prod_{j \in A} \bar{\theta}_j \theta_j \biggr] \biggl[ \prod_{k \in B} (1 - \bar{\theta}_k \theta_k ) \biggr] e^{\theta^{\dagger} \Lambda \theta} , \notag \\
=& \; 
\frac{1}{\det \Lambda} \int \biggl[ \prod_{i=1}^p d\theta_i d\bar{\theta}_i \biggr] \biggl[ \prod_{j \in A} \delta(\bar{\theta}_j)  \delta(\theta_j) \biggr] \biggl[ e^{- \theta_B^{\dagger} \theta_B } \biggr] e^{\theta^{\dagger} \Lambda \theta} , \notag \\
=& \; \frac{1}{\det \Lambda} \det( \Lambda_{BB} - I ) , \label{eq: joint_grassmann}
\end{align}
where $I$ denotes the identity matrix.
The above equation indicates that the joint probabilities are expressed as principal minors of the matrix $\Lambda - I$ divided by $\det \Lambda$.

Next, we turn to the marginal distribution.
We write the index labels of the marginalized and remaining variables as $M$ and $R$, as defined by Sec.~\ref{sec: result}.
Then, by completing the square, the marginalization of the Grassmann variable is calculated as
\begin{align}
\frac{1}{Z_{x_R}} \; e^{H(\bar{\theta}_R, \theta_R)} \equiv & \;  \frac{1}{\det \Lambda} \int d\theta_M d\bar{\theta}_M \, \exp \Bigl\{\theta^{\dagger} \Lambda \theta \Bigr\}, \notag \\
=& \;  \frac{1}{\det \Lambda} \int d \theta_M d\bar{\theta}_M
\exp \Bigl\{ \theta_R^{\dagger} ( \Lambda_{RR} - \Lambda_{RM} \Lambda_{MM}^{-1} \Lambda_{MR} ) \theta_R \notag \\
 & \hspace{4cm} + (\theta_M^{\dagger}  + \theta_R^{\dagger} \Lambda_{RM} \Lambda_{MM}^{-1}) \Lambda_{MM} \bigl( \theta_M + \Lambda_{MM}^{-1} \Lambda_{MR} \theta_R \bigr) \Bigr\}, \notag \\
=&\; \frac{1}{\det \Lambda_{R|M} } \exp \Bigl\{ \theta_R^{\dagger} (\Lambda_{RR} - \Lambda_{RM} \Lambda_{MM}^{-1} \Lambda_{MR} ) \theta_R \Bigr\}, \notag \\
=& \; \det \Sigma_{RR}  \; \exp \Bigl\{\theta_R^{\dagger} \Sigma_{RR}^{-1} \theta_R \Bigr\} ,
\end{align}
where
\begin{align}
\Sigma_{RR} =& \; \Lambda_{R|M}^{-1}, \notag \\
=& \; \bigl[ \Lambda_{RR} - \Lambda_{RM} \Lambda_{MM}^{-1} \Lambda_{MR} \bigr]^{-1}. 
\end{align}
The parameter of the marginal Hamiltonian is just a principal submatrix of $\Sigma$ with the same indices of rows and columns.
That is, the diagonal and off-diagonal elements of the matrix $\Sigma$ denote the mean and the covariance with all the other variables marginalized out.
When the product of the off-diagonal elements $-\Sigma_{ij} \Sigma_{ji}$ vanishes, the variables $x_i$ and $x_j$ are unconditionally independent or marginally independent.
Higher-order central moments can also be calculated by the Grassmann integral.
For example, the central moment for the variables with the index label $R$, Eq.~(\ref{eq: central_moment}), is derived by
\begin{align}
m_{R} \equiv & \, \mathrm{E} \biggl[ \; \prod_{i \in R} (x_i - \mu_i ) \biggr], \notag \\
=& \, \frac{1}{\det \Lambda} \int d\theta_P d \bar{\theta}_P \prod_{i \in R} ( \bar{\theta}_i \theta_i - \mu_i ) \, e^{\theta^{\dagger} \Lambda \theta}, \notag \\
= & \, \det \Bigl[ \Sigma_{RR} - \mathrm{diag}(\mu_R) \Bigr],
\end{align}
where $\mathrm{diag}(\mu_R)$ is the diagonal matrix defined by Eq.~(\ref{eq: kronecker}):
\begin{align}
\mathrm{diag}(\mu_R) \equiv & \; \delta_{ij} \, \mu_i, \hspace{0.5cm} (i, j \in R).
\end{align}

Then, we discuss the conditional distribution.
We define the index labels for the variables as defined by Sec.~\ref{sec: result}.
That is, we write index labels for the variables observed as $x_i=1$ and $x_i=0$ as $A$ and $B$ and write these variables as $x_A$ and $x_B$, respectively.
The remaining indices after conditioning are represented by the set difference by these indices $R = P \setminus (A \cup B) \equiv P \setminus C$.
Then, the conditional Hamiltonian $H(\bar{\theta}_R, \theta_R | x_C)$ is given by
\begin{align}
& \frac{1}{Z_{x_R | x_C}} \; e^{H(\bar{\theta}_R, \theta_R | x_C)} \notag \\ 
=&  \;  \frac{1}{p(x_A=1, x_B=0)} \;  \frac{1}{\det \Lambda} \int d\theta_C d\bar{\theta}_C \, (\bar{\theta}_A \theta_A) (1 - \bar{\theta}_B \theta_B) \, \exp \Bigl\{ \theta^{\dagger} \Lambda \theta \Bigr\},  \notag \\
=& \frac{1}{\det (\Lambda_{BB} - I) \det ( \Lambda_{RR} - \Lambda_{RB} (\Lambda_{BB} - I)^{-1} \Lambda_{BR} ) }
\int d \theta_C d \bar{\theta}_C \; \delta(\bar{\theta}_A) \delta(\theta_A) \exp\Bigl\{ - \theta_B^{\dagger} \theta_B
+ \theta^{\dagger} \Lambda \theta \Bigr\}, \notag \\
=&  \frac{1}{\det (\Lambda_{BB} - I) \det ( \Lambda_{RR} - \Lambda_{RB} (\Lambda_{BB} - I)^{-1} \Lambda_{BR} ) } \notag \\
& \int d\theta_B d \bar{\theta}_B \exp\Bigl\{ \theta_R^{\dagger} \bigl[ \Lambda_{RR} - \Lambda_{RB} (\Lambda_{BB} - I)^{-1} \Lambda_{BR} \bigr] \theta_R \notag \\
& \hspace{3cm} + \bigl[ \theta_{B}^{\dagger} + \theta_R^{\dagger} \Lambda_{RB} (\Lambda_{BB} - I)^{-1} \bigr]
(\Lambda_{BB} - I) \bigl[ \theta_{B} + (\Lambda_{BB} - I)^{-1} \Lambda_{BR} \theta_{R} \bigr] \Bigr\}, \notag \\
=& \frac{1}{\det(\Lambda_{RR} - \Lambda_{RB}(\Lambda_{BB} - I)^{-1} \Lambda_{BR})} \exp\Bigl\{ \theta_{R}^{\dagger} \bigl[ \Lambda_{RR} - \Lambda_{RB} (\Lambda_{BB} - I)^{-1} \Lambda_{BR} \bigr] \theta_R \Bigr\}, \notag \\
=&\;  \det \Sigma_{R | x_C}  \, \exp \Bigl\{ \theta_R^{\dagger} \, \Sigma_{R|x_C}^{-1} \theta_R \Bigr\},
\end{align}
where
\begin{align}
\Sigma_{R | x_C} =& \;  \Sigma_{RR} - \Sigma_{RC} \bigl[ \Sigma_{CC} - \mathrm{diag}(1 - x_C) \bigr]^{-1} \Sigma_{CR}  \notag \\
=& \; \bigl[ \Lambda_{RR} - \Lambda_{RB} (  \Lambda_{BB} - I  )^{-1} \Lambda_{BR} \bigr]^{-1}
\end{align}
can be read from Eq.~(\ref{eq: lambda_sigma_tilde}).
The matrix $\Sigma_{R|x_C}$ can be rewritten by the Schur complement of the following matrix $\tilde{\Sigma}$ with respect to the principal submatrix $\tilde{\Sigma}_{CC}$,
\begin{align}
\Sigma_{R | x_C} = & \; \tilde{\Sigma}_{R | C} , \notag \\
= & \;  \Sigma_{RR} - \tilde{\Sigma}_{RC} \tilde{\Sigma}_{CC}^{-1} \Sigma_{CR} ,
\end{align}
where
\begin{align}
\tilde{\Sigma} \equiv & \begin{bmatrix} \Sigma_{RR} & \tilde{\Sigma}_{RC} \\ \Sigma_{CR} & \tilde{\Sigma}_{CC} \end{bmatrix}
\equiv  \begin{bmatrix} 
\Sigma_{RR} & \Sigma_{RA} & -\Sigma_{RB} \\
\Sigma_{AR} & \Sigma_{AA} & - \Sigma_{AB} \\
\Sigma_{BR} & \Sigma_{BA} & I - \Sigma_{BB} \end{bmatrix}
\equiv  \tilde{\Lambda}^{-1}. \label{eq: tilde_conditional}
\end{align}
The matrix $\tilde{\Sigma}$ corresponds to the original matrix $\Sigma$ with the mean $\Sigma_{BB}$ and sign of the covariance parameters $( \Sigma_{RB}, \Sigma_{AB})$ for the variables $x_B$ inverted as $ I - \Sigma_{BB}$ and $(- \Sigma_{RB}, -\Sigma_{AB})$, respectively.
In other words, observing the dummy variable $x_i$ as $x_i = 0$ is equivalent to observing the dummy variable $\tilde{x}_i \equiv 1 - x_i$ with the dummy coding inverted as $\tilde{x}_i = 1$.
Therefore, our formalism is a symmetric formalism that does not depend on how to encode binary variables.

The matrix $\Lambda$ can also be interpreted intuitively.
We consider the case that the conditioning variables are all observed as $x_i = 1$.
Then, we see that the diagonal element of $\Lambda$ represents the reciprocal of the mean of the dummy variable conditioned on all the other variables observed as $x_A = 1$.
Furthermore, the off-diagonal elements can be interpreted as the partial correlation, similarly to the multivariate Gaussian distribution.
To see this, we consider the conditional distribution of $x_R =(x_i, x_j) $ given $x_A=1, \; A = P \setminus R $.
The corresponding conditional Hamiltonian is given by
\begin{align}
\frac{1}{Z_{x_R | x_A=1}} \; e^{H(\bar{\theta}_R, \theta_R | x_A = 1)} =& \;  \frac{1}{\det \Lambda_{RR}} \, e^{\theta_R^{\dagger} \Lambda_{RR} \theta_R} ,\notag \\
=& \; \det \Sigma_{R | A} \; e^{\theta_R^{\dagger} \Sigma_{R|A}^{-1} \theta_R}, 
\end{align}
where
\begin{align}
\Sigma_{R |A} =& \begin{bmatrix} \Sigma_{ii|A} & \Sigma_{ij|A} \\ \Sigma_{ji|A} & \Sigma_{jj |A} \end{bmatrix} 
= \Lambda_{RR}^{-1}
= \frac{1}{\det \Lambda_{RR} } \begin{bmatrix} \Lambda_{jj} & - \Lambda_{ij} \\ - \Lambda_{ji} & \Lambda_{ii} \end{bmatrix}.
\end{align}
Then the correlation between $x_i$ and $x_j$ for the conditional distribution, i.e., the partial correlation $\rho_{ij|A}$, is expressed as the product of the off-diagonal elements of $\Lambda_{RR}$:
\begin{align}
\rho_{ij|A} =& \; \frac{ - \Sigma_{ij|A} \Sigma_{ji|A} }{\sqrt{ \Sigma_{ii|A} ( 1- \Sigma_{ii|A} ) \Sigma_{jj|A} (1 - \Sigma_{jj|A} )}}, \notag \\
=& \; \frac{ - \Lambda_{ij} \Lambda_{ji} }{ \sqrt{ \Lambda_{jj} (\det \Lambda_{RR} - \Lambda_{jj} ) \Lambda_{ii} (\det \Lambda_{RR} - \Lambda_{ii} )}}. \label{eq: partial_correlation}
\end{align}
Therefore, the off-diagonal elements of $\Lambda$ can be interpreted as the partial correlation with all the other variables observed as $x_A=1$.
The partial correlation for the general conditioning $x_C = (x_A=1, x_B=0)$ other than $x_A=1$ can also be interpreted in the same way.
In this case, we first define the matrix $\tilde{\Sigma}$ in which the dummy coding of the variables observed as $x_B=0$ is inverted to $\tilde{x}_B=1$ as in Eq.~(\ref{eq: tilde_conditional}).
Then, the product of the off-diagonal elements of its inverse matrix $\tilde{\Lambda} = \tilde{\Sigma}^{-1}$ represents the magnitude of the partial correlation under that conditioning.
The partial correlation for the general conditioning is given by the same expression, Eq.~(\ref{eq: partial_correlation}), except that $\Lambda$ is replaced by $\tilde{\Lambda}$.

Last, we should mention normalization and the positivity of our probability distribution.
Since the analytical expression for the partition function is obtained in our formalism, normalization for the joint distribution can be checked analytically.
The joint probabilities of $p$-dimensional binary variables are expressed as principal minors of the matrix $\Lambda - I$ divided by $\det \Lambda$ as shown in Eq.~(\ref{eq: joint_grassmann}).
When we notice the identity regarding the summation over all principal minors,
\begin{align}
\sum_{B \subseteq P } \det(\Lambda_{BB} - I) = \det \Lambda,
\end{align}
we see that normalization of the joint distribution is satisfied by definition.
On the other hand, the property that all joint probabilities are greater than or equal to zero does not necessarily hold true in general.
Expressed in the terminology of linear algebra, the property that all joint probabilities, i.e., all principal minors of $\Lambda - I$, must be greater than or equal to zero is equivalent to that the matrix $\Lambda - I$ must be a $P_0$ matrix, which is an important property in various applications~\cite{Berman1994}.
When the matrix $\Lambda - I$ is a $P$ matrix, the matrices $\Sigma$ and $I - \Sigma$ are also $P$ matrices.
It is because, if the matrix $A = \Lambda -I = \Sigma^{-1} - I $ is a $P$ matrix, the matrices $I + F_A = 2 \Sigma$ and $I - F_A = 2 (I - \Sigma)$ are also $P$ matrices from the theorem on $P$ matrices~\cite{Tsatsomeros2002}, where $F_A = (I - A)(I + A)^{-1} = 2 \Sigma - I$ is the Cayley transform of $A = \Lambda -I$.
When the matrix $\Lambda - I$ is a $P$ matrix, the positivity of the marginal and conditional distributions can also be confirmed in terms of linear algebra.
The marginal probabilities are expressed as all the principal minors of the Schur complements of the matrix
\begin{align}
\Lambda - I + \mathrm{diag}(\delta_{PM}) \equiv \begin{bmatrix} \Lambda_{RR} - I & \Lambda_{RM} \\ \Lambda_{MR} & \Lambda_{MM} \end{bmatrix}.
\end{align}
with respect to $\Lambda_{MM}$.
Here, the matrix $\Lambda - I + \mathrm{diag}(\delta_{PM}) $ is still a $P$ matrix since adding a diagonal matrix with nonnegative diagonal elements does not change the positivity of all principal minors~\cite{Tsatsomeros2002}.
Since the Schur complement of a $P$ matrix is also a $P$ matrix, it follows that all the principal minors, i.e., all the marginal probabilities, are positive
For the conditional probabilities, their positivity is rephrased as the $R' \, (\subseteq R)$ principal minors of the Schur complement of $\Lambda - I$ with respect to $\Lambda_{BB} - I$.
Again, since the Schur complement of a $P$ matrix is a $P$ matrix, all of the conditional probabilities are positive.
Normalization and the positivity of our probability distribution are in contrast to those of the conventional multivariate Bernoulli distribution.
In the multivariate Bernoulli distribution, the partition function is not given analytically but has to be summed numerically over all possible states.
On the other hand, the property that all joint probabilities are positive is satisfied by definition because probability distributions are given by the exponential function of the polynomial in the dummy variables as shown in Eq.~(\ref{eq: Bernoulli_distribution}).

\section{Parameter estimation \label{sec: parameter_estimation} }
In our model, we have to resort to numerical computation to estimate model parameters from observed data.
Hence, in this section, we discuss parameter estimation and the sampling distribution of the parameter estimates.
Below, we define index labels for the variables used in this section.
We denote the set of all indices for $p$-dimensional binary variables as $P \equiv \{ 1, 2, \dots, p\}$.
Then, we write the index labels for the variables observed as $x_i=1$ and $x_i=0$ as $A$ and $B$, and write these variables as $x_A$ and $x_B$, respectively.
We denote a specific realization of the dummy vector as $\delta$, for example, $\delta = (1,0,1,1,0)$ for five-dimensional variables.
Generated data are denoted by $\mathcal{D} = \{x_{1P}, x_{2P}, \dots, x_{NP} \}$, where $x_{nP}, \;  n \in \{ 1, 2, \dots, N\} $, is a $p$-dimensional vector of dummy variables.

\subsection{Diagonally dominant parametrization}
A common method of parameter estimation is maximum likelihood estimation given observed data $\mathcal{D}$.
In our model, the log-likelihood function is expressed as
\begin{align}
l(\Sigma | \mathcal{D} )
=&  \sum_{n=1}^N \log p(x_{nP} | \Sigma ), \notag \\
= & \, N   \sum_{\delta}  \frac{n_{\delta} }{N} \log p(x_P = \delta | \Sigma ), \notag \\
\equiv & \,  N \sum_{\delta} q_\delta \log \pi_\delta(\Sigma) , \label{eq: cross_entropy}
\end{align}
where $n_{\delta}$ is the number of times we observed the state as $x_{P}= \delta$, which satisfies $\sum_{\delta} n_{\delta} = N$.
In other words, the log-likelihood is expressed as the cross entropy between the empirical joint distribution $q_{\delta}$ and the distribution by the model $\pi_{\delta}(\Sigma) \equiv p(x_P=\delta | \Sigma)$.
Since it is difficult to solve the log-likelihood maximization analytically, one has to resort to numerical calculations.

In order for our model to make sense as a probability distribution, all the joint probabilities by the model $\pi_{\delta}(\Sigma)$ must take non-negative values.
This is equivalent to requiring $\Lambda-I$ to be a $P_0$ matrix.
When all $q_{\delta}$ are nonzero, iterative maximization of the log-likelihood ensures naturally that the probabilities $\pi_{\delta}(\Sigma)$ remain positive as long as the initial parameters satisfy the positivity condition, $\pi_{\delta}(\Sigma) > 0 \; (\forall \, \delta)$.
Indeed, the log-likelihood tends to $-\infty$ whenever one of the probabilities $\pi_{\delta}(\Sigma)$ approaches to zero.
However, since the sample size is finite in practice, some of the empirical joint probabilities $q_{\delta}$ become zero.
Then, the difficulty arises that the corresponding joint probability by the model $\pi_{\delta}(\Sigma)$ can take a negative value during parameter estimation.
The method of introducing a prior distribution for observing all possible states, i.e., pseudocounts, works as a regularization, but this procedure is not practically available when the dimension of the variable is large, because the number of states that have to be summed in the log-likelihood increases exponentially.
Hence, we have to truly enforce that $\Lambda-I$ be a $P$ or $P_0$ matrix.

One solution to the positivity of a probability distribution is to parametrize $\Lambda$ using strictly row diagonally dominant matrices $B$ and $C$ with positive diagonal entries:
\begin{align}
\Lambda = B C^{-1} + I,
\end{align}
where each element of $B$ and $C$ satisfies
\begin{align}
b_{ii} > & \sum_{j \neq i} |b_{ij}|, \\
c_{ii} > & \sum_{j \neq i} |c_{ij}|.
\end{align}
This parametrization enforces $\Lambda - I$ to be a $P$ matrix~\cite{Tsatsomeros2002}.

In our model, the number of terms in the log-likelihood is proportional to the number of unique states that were observed, which provides a computational advantage over the multivariate Bernoulli distribution.
In the multivariate Bernoulli distribution, maximum likelihood estimation requires summing the probabilities over all possible states, which increases exponentially as the dimension of the variables increases.
In our model, however, the positivity of the probability distribution is not ensured by definition.
The pseudocounts work as a regularization, but this procedure is not practically available, due to computational complexity as in the case of the multivariate Bernoulli distribution.
The diagonally dominant parametrization allows us to take advantage of the computational complexity of our model.
In this parametrization, the computational complexity of maximum likelihood estimation becomes proportional to the number of unique observed states, not to the number of all possible states as is required in the case of the multivariate Bernoulli distribution.

\subsection{Choosing parameters for random number generator \label{sec: maximum_entropy} }
In this subsection, we discuss how to choose the model parameters, based on which we generate the synthetic dataset.
Since the covariance structure of our model is determined by the product of off-diagonal elements of the matrix, $-\Sigma_{ij} \Sigma_{ji}$, the parameters $\Sigma_{ij}$ themselves are not fixed uniquely even though we set the mean and the covariance by hand.
That is, there are multiple parameter matrices that can lead to exactly the same mean and covariance, but generate different distributions.
Hence, we choose the most natural model, in the sense that has the least number of assumptions, subject to the constraint of having a specified mean and covariance.
The remaining parameters of the model $\Sigma_{ij} \; (i<j)$ with a specified mean and covariance are determined by maximizing the entropy of the joint distribution $\mathbb{H}(p | \Sigma_{ij})$,
\begin{align}
\mathbb{H}(p | \Sigma_{ij}) = - \sum_{\delta} p(x_P=\delta | \Sigma_{ij} ) \log p(x_P=\delta | \Sigma_{ij}),
\end{align}
where the summation has been taken over all possible states.
Since it is difficult to find the global maximum of the above entropy function, we repeated the estimation with different initial values and regarded the parameter that had the largest entropy and was estimated multiple times as that of the global maximum.
The maximum entropy principle requires evaluating the probabilities for all possible states, so it is not practically available when the dimension of the variables is large, because the number of states increases exponentially.

\subsection{Sampling distribution of maximum likelihood estimates}
We generated synthetic datasets using the parameters chosen by the maximum entropy principle and investigate the sampling distribution of maximum likelihood estimates.
The dimension of the variables used to generate the synthetic datasets is $p=5$, and the mean parameters are $\mu_1=0.85$, $\mu_2=0.46$, $\mu_3=0.74$, $\mu_4=0.70$, $\mu_5=0.80$.
The correlation coefficients of the variables are $\rho_{12}=-0.21$, $\rho_{13}=-0.07$, $\rho_{14}=-0.01$, $\rho_{15} =0.01$, $\rho_{23} = 0.11$, $\rho_{24}=-0.03$, $\rho_{25} = 0.04$, $\rho_{34}=0.44$, $\rho_{35}=-0.07$, $\rho_{45} =0.37$, where we have defined the correlation coefficient for dummy variables by the Pearson correlation coefficient, as defined by Eq.~(\ref{eq: pearson_correlation}).
The given mean and covariances were chosen by a uniform distribution.
The parameters used to generate the synthetic dataset are shown in Eq.~(\ref{eq: maximum_solution}).

\begin{figure}[htbp]
\centering
\includegraphics[width=14cm]{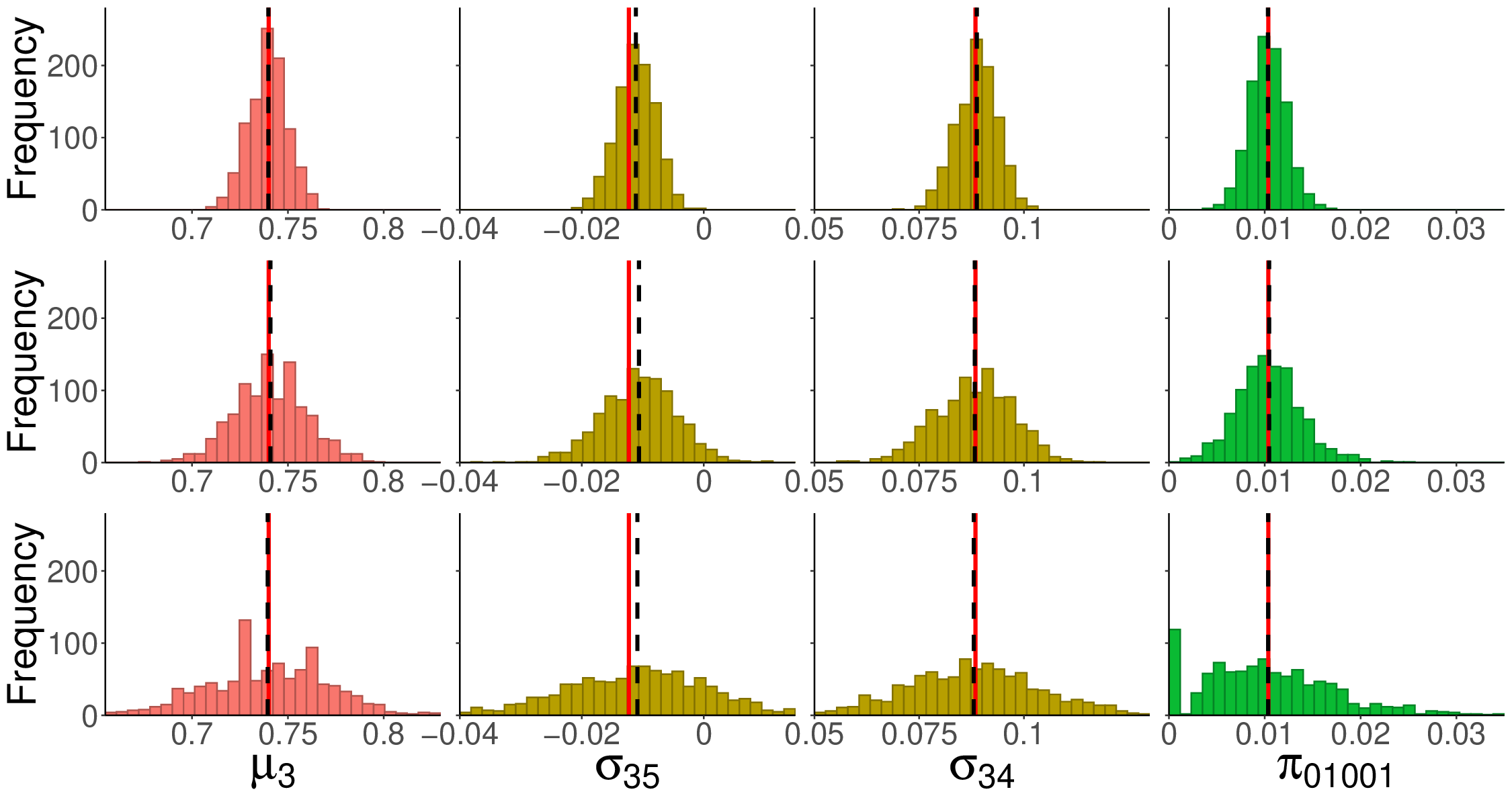}
\caption{Examples of the sampling distributions of the maximum likelihood estimates for the mean parameter $\mu_3$, covariance parameters $\sigma_{35}$, $\sigma_{34}$, and the joint probability by the estimated model $\pi(x_P=(0,1,0,0,1))$.
The first, second and third rows correspond to the results of the sample sizes $N=2000$, $N=600$, and $N=200$, respectively.
The trial size, i.e., the number of times the estimates are computed, is $M=1000$.
The red solid lines denote the true values and the black dashed lines denote the mean values of the sampling distributions.
}
\label{fig: sampling_distribution_estimates}
\end{figure}

Figure~\ref{fig: sampling_distribution_estimates} shows the sampling distributions of the maximum likelihood estimates for the mean and covariance parameters and the joint distribution by the estimated model for different sample sizes $N$.
We see that the maximum likelihood estimates of the mean agree with the empirical mean.
As in the case of statistics, we observe that each estimate converges to the true values of the parameters asymptotically as the sample size goes to infinity.
Although the sampling distribution can be skewed when the sample size is small, it becomes asymptotically normal as the sample size increases.
In other words, the maximum likelihood estimator appears to be consistent and asymptotically normal.
The standard errors of the sampling distributions decrease as $1/\sqrt{N}$ as the sample size increases.
These results suggest that the usual statistical inference on a model parameter such as hypothesis testing and confidence interval estimation is available.

\begin{figure}[htbp]
\centering
\includegraphics[width=14cm]{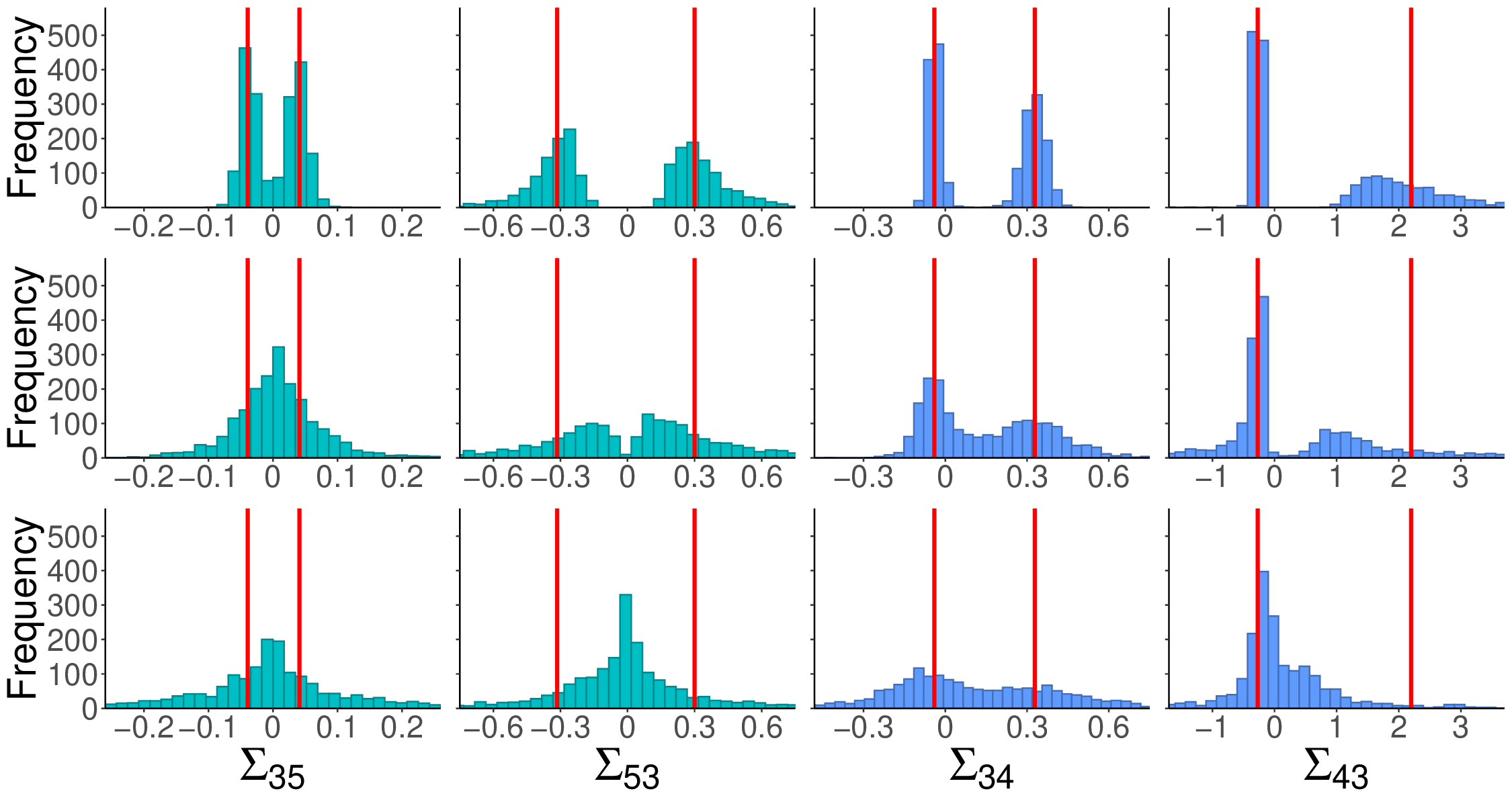}
\caption{Examples of the sampling distributions of the maximum likelihood estimates for the parameters $\Sigma_{ij}$.
The first, second, and third rows correspond to the results of the sample sizes $N=20\,000$, $N=2000$, and $N=200$, respectively.
The trial size is $M=1000$.
The red solid lines denote the value of the true parameters.
Due to the degrees of freedom of the matrix transposition, there exist two lines corresponding to the true parameters.
}
\label{fig: sampling_distribution_estimates_sigma}
\end{figure}

Figure~\ref{fig: sampling_distribution_estimates_sigma} shows the sampling distributions of the maximum likelihood estimates for the parameters $\Sigma_{ij}$ for different sample sizes $N$.
As discussed in Sec.~\ref{sec: property}, the parameter matrices $\Sigma$ and $\Lambda$ have degrees of freedom as shown in Eq.~(\ref{eq: dilatation}) and (\ref{eq: transposition}).
These degrees of freedom can also be read from the expression for the Grassmann integral.
To compare the estimated parameters with the true parameters, we fixed the degrees of freedom of model parameters.
Assuming $\Sigma_{i1} \;  (i \neq 1)$ to be nonzero, we fix the degrees of freedom of the constant multiplication so that $\Sigma_{i1} = -1$.
That is, the parameters $\Sigma_{1j} \; (j \neq 1)$ simply represent the covariances with the dummy variable $x_1$.
However, we failed to find a natural way to fix the degree of freedom for the matrix transposition.
Hence, in Fig.~\ref{fig: sampling_distribution_estimates_sigma}, we show both of the estimated parameters corresponding to the degrees of freedom for the matrix transposition.
Although the sampling distribution of the estimate $\Sigma_{ij}$ has large dispersion and is unimodal when the sample size is small, the estimates converge to one of the true parameters asymptotically.
In addition, we empirically observe that the sampling distribution becomes asymptotically normal.
In other words, the maximum likelihood estimator appears to be consistent and asymptotically normal.

\section{Conclusion \label{sec: conclusion} }
We formulated a probability distribution for multivariate binary variables.
The partition function, central moments, and the joint, marginal, and conditional distributions are analytically expressed in terms of the parameter matrix $\Sigma$, which is a matrix analogous to the covariance matrix in the multivariate Gaussian distribution.
The proposed model has many similarities to the multivariate Gaussian distribution.
For example, a principal submatrix of $\Sigma$ describes the marginal distribution, the diagonal elements of which represent the mean, and the products of the off-diagonal elements represent the covariance.
Its inverse matrix $\Lambda = \Sigma^{-1}$ describes the conditional distribution; the diagonal elements represent the reciprocal of the mean and the off-diagonal elements can be interpreted as a kind of partial correlation conditioned on all the other variables.
The uncorrelatedness of variables for the marginal and conditional distributions is equivalent to unconditional and conditional independence, respectively.
Furthermore, the joint probabilities corresponding to all $2^p$ possible states for a $p$-dimensional binary variable are expressed as all possible principal minors of the matrix $\Lambda - I$.
The property that all joint probabilities must be greater than or equal to zero can be rephrased in the terminology of linear algebra as that the matrix $\Lambda - I$ must be a $P_0$ matrix.
These properties are in contrast to those of the conventional multivariate Bernoulli distribution, where the joint probabilities of all states are always greater than zero because they are expressed as an exponential function of a polynomial in the dummy variables.

In our model, we have to resort to numerical computation to estimate model parameters from observed data.
We discussed maximum likelihood estimation of model parameters.
Since it is not ensured that the probability distribution is nonnegative by definition, we parametrize the model parameters $\Lambda - I$ using strictly diagonally dominant matrices, which enforces the probabilities to be positive.
Owing to this diagonally dominant parametrization, we only have to evaluate the joint probabilities for unique observed states in maximum likelihood estimation, which is in contrast to the conventional multivariate Bernoulli distribution where one has to evaluate the probabilities for all possible states.
In other words, our method does not suffer from the computational complexity of the conventional multivariate Bernoulli distribution in practice.

Since we have analytical expressions for the marginal and conditional distributions, we can easily generate random numbers for correlated binary variables by repeating Bernoulli trials.
We investigated the sampling distributions of various statistics and maximum likelihood estimates by using synthetic datasets.
We numerically confirm that the sampling distributions of the statistics are consistent with the theoretical predictions.
We empirically observed that the maximum likelihood estimator appears to be consistent and asymptotically normal.

It is important to demonstrate that the proposed method can fit real data well.
Another direction is to explore theoretical properties of the sampling distribution of an estimated parameter to make a statistical inference on a model parameter such as hypothesis testing and confidence interval estimation.
Since our method has many similarities to the multivariate Gaussian distribution, it has many potential applications, where the covariance structure of random variables is extensively utilized.
Examples include the hierarchical and nonhierarchical clustering, such as $k$-means clustering and the mixture distribution model, and anomaly detection like the Hotelling's $T^2$ method.
There, a method similar to the multivariate Gaussian distribution will be extended and applied to binary random variables.
Our method in turn will also be useful in studying the behavior of gases or magnets in statistical physics, which have conventionally been analyzed using the Ising model.


\appendix

\section{Formulas for a partitioned matrix \label{sec: matrix_identity} }
In this appendix, we describe the formulas of a partitioned matrix in our notation.
For a partitioned matrix with subsets index $A$ and $B$,
\begin{align}
\Sigma = \begin{bmatrix} \Sigma_{AA} & \Sigma_{AB} \\ \Sigma_{BA} & \Sigma_{BB} \end{bmatrix},
\end{align}
where $\Sigma_{BB}$ is invertible, the Schur complement of $\Sigma$ with respect to $\Sigma_{BB}$ is defined by
\begin{align}
\Sigma / \Sigma_{BB} = & \Sigma_{A|B}, \\
\equiv &  \Sigma_{AA} - \Sigma_{AB} \Sigma_{BB}^{-1} \Sigma_{BA} .
\end{align}
If the diagonal blocks $\Sigma_{AA}$ and $\Sigma_{BB}$ are both invertible, then the inverse of the partitioned matrix is expressed in the following two ways:
\begin{align}
\begin{bmatrix} \Sigma_{AA} & \Sigma_{AB} \\ \Sigma_{BA} & \Sigma_{BB} \end{bmatrix}^{-1}
=& \begin{bmatrix}
\Sigma_{AA}^{-1} + \Sigma_{AA}^{-1} \Sigma_{AB} \Sigma_{B|A}^{-1} \Sigma_{BA} \Sigma_{AA}^{-1} & - \Sigma_{AA}^{-1} \Sigma_{AB} \Sigma_{B|A}^{-1} \\
- \Sigma_{B|A}^{-1} \Sigma_{BA} \Sigma_{AA}^{-1} & \Sigma_{B|A}^{-1} 
\end{bmatrix}, \notag  \\
=& \begin{bmatrix}
\Sigma_{A|B}^{-1} & -\Sigma_{A|B}^{-1} \Sigma_{AB} \Sigma_{BB}^{-1} \\
- \Sigma_{BB}^{-1} \Sigma_{BA} \Sigma_{A|B}^{-1} & \Sigma_{BB}^{-1} + \Sigma_{BB}^{-1} \Sigma_{BA} \Sigma_{A|B}^{-1} \Sigma_{AB} \Sigma_{BB}^{-1}
\end{bmatrix}
=\begin{bmatrix} \Lambda_{AA} & \Lambda_{AB} \\ \Lambda_{BA} & \Lambda_{BB} \end{bmatrix}, 
\end{align}
which is known as the partitioned inverse formula and can be derived using elementary operations for a partitioned matrix.
From the above equations, we can read the Woodbury matrix identity or the matrix inversion lemma,
\begin{align}
\bigl[ \Sigma_{AA} - \Sigma_{AB} \Sigma_{BB}^{-1} \Sigma_{BA} \bigr]^{-1} =
 \Sigma_{AA}^{-1} + \Sigma_{AA}^{-1} \Sigma_{AB} \Sigma_{B|A}^{-1} \Sigma_{BA} \Sigma_{AA}^{-1}.
\end{align}
Furthermore, we can read the relation between $\Sigma_{AB}$ and $\Lambda_{AB}$:
\begin{align}
\Sigma_{AB} \Sigma_{BB}^{-1} =& - \Lambda_{AA}^{-1} \Lambda_{AB}, \label{eq: Sigma_AB_left} \\
\Sigma_{AA}^{-1} \Sigma_{AB} =& - \Lambda_{AB} \Lambda_{BB}^{-1}. \label{eq: Sigma_AB_right}
\end{align}
In our notation, the matrix determinant lemma is given by
\begin{align}
\det \Sigma = \det \Sigma_{AA} \det \Sigma_{B|A} = \det \Sigma_{BB} \det \Sigma_{A|B}. \label{eq: matrix_determinant_lemma}
\end{align}

The inverse of the matrix $\tilde{\Sigma}$,
\begin{align}
\tilde{\Sigma} = \begin{bmatrix} \Sigma_{AA} & -\Sigma_{AB} \\ \Sigma_{BA} & I - \Sigma_{BB} \end{bmatrix},
\end{align}
 can be expressed in terms of $\Lambda$.
In fact, from the partitioned inverse formula, the inverse of the matrix $\tilde{\Sigma}$ is represented by
\begin{align}
& \tilde{\Sigma}^{-1} \notag \\
=& \begin{bmatrix}
\Sigma_{AA}^{-1} - \Sigma_{AA}^{-1} \Sigma_{AB} \bigl[ I - \Sigma_{BB} + \Sigma_{BA} \Sigma_{AA}^{-1} \Sigma_{AB} \bigr]^{-1} \Sigma_{BA} \Sigma_{AA}^{-1} & \Sigma_{AA}^{-1}  \Sigma_{AB}  \bigl[ I - \Sigma_{BB} + \Sigma_{BA} \Sigma_{AA}^{-1} \Sigma_{AB} \bigr]^{-1} \notag \\
-  \bigl[ I - \Sigma_{BB} + \Sigma_{BA} \Sigma_{AA}^{-1} \Sigma_{AB} \bigr]^{-1} \Sigma_{BA} \Sigma_{AA}^{-1} &  \bigl[ I - \Sigma_{BB} + \Sigma_{BA} \Sigma_{AA}^{-1} \Sigma_{AB} \bigr]^{-1} \end{bmatrix}, \\
=& \begin{bmatrix}
\Sigma_{AA}^{-1} - \Sigma_{AA}^{-1} \Sigma_{AB} \Lambda_{BB} ( \Lambda_{BB} - I)^{-1} \Sigma_{BA} \Sigma_{AA}^{-1} & \Sigma_{AA}^{-1} \Sigma_{AB} \Lambda_{BB} (\Lambda_{BB} - I )^{-1} \\
- \Lambda_{BB} (\Lambda_{BB} - I )^{-1} \Sigma_{BA} \Sigma_{AA}^{-1} & \Lambda_{BB} (\Lambda_{BB} - I)^{-1}
\end{bmatrix},
\end{align}
where we have used the relation,
\begin{align}
\bigl[ I - \Sigma_{BB} + \Sigma_{BA} \Sigma_{AA}^{-1} \Sigma_{AB}  \bigr]^{-1} =& \bigl[ I - \Sigma_{B|A} \bigr]^{-1}, \notag \\
=&  \frac{\Lambda_{BB}}{\Lambda_{BB} - I}.
\end{align}
Using the relation between $\Sigma_{AB}$ and $\Lambda_{AB}$, Eqs.~(\ref{eq: Sigma_AB_left}) and (\ref{eq: Sigma_AB_right}), we arrive at
\begin{align}
& \tilde{\Sigma}^{-1} \notag \\
=& \begin{bmatrix}
\Sigma_{AA}^{-1} - \Sigma_{AA}^{-1} \Sigma_{AB} \bigl[ - \Lambda_{BB} + \Lambda_{BB} + \Lambda_{BB}(\Lambda_{BB} - I)^{-1} \bigr] \Sigma_{BA} \Sigma_{AA}^{-1} & - \Lambda_{AB} (\Lambda_{BB} - I)^{-1} \notag \\
 (\Lambda_{BB} - I)^{-1} \Lambda_{BA} & \Lambda_{BB} (\Lambda_{BB} - I)^{-1} 
\end{bmatrix} , \\
=& \begin{bmatrix}
\Lambda_{AA} - \Lambda_{AB} (\Lambda_{BB} - I)^{-1} \Lambda_{BA} & - \Lambda_{AB} (\Lambda_{BB} - I)^{-1} \\
 (\Lambda_{BB} - I)^{-1} \Lambda_{BA} & \Lambda_{BB} (\Lambda_{BB} - I)^{-1} 
\end{bmatrix}. \label{eq: flipped_formula_derivation}
\end{align}
From the diagonal block with indices $R \subseteq A$ of the above expression and the corresponding partitioned inverse formula for $\tilde{\Sigma}$, we see the relation,
\begin{align}
\Lambda_{RR} - \Lambda_{RB} (\Lambda_{BB} - I)^{-1} \Lambda_{BR} =& \tilde{\Sigma}_{R|C}, \notag \\
=& \tilde{\Sigma}_{RR} - \tilde{\Sigma}_{RC} \tilde{\Sigma}_{CC}^{-1} \tilde{\Sigma}_{CR}, \notag \\
=& \Sigma_{RR} - \begin{bmatrix} \Sigma_{RA'} & - \Sigma_{RB} \end{bmatrix}
\begin{bmatrix} \Sigma_{A'A'} & - \Sigma_{A' B} \\ \Sigma_{BA'} & I - \Sigma_{BB} \end{bmatrix}^{-1}
\begin{bmatrix} \Sigma_{A' R} \\ \Sigma_{BR} \end{bmatrix}, \label{eq: lambda_sigma_tilde}
\end{align}
where $A' = A \setminus R$ and $C = A' \cup B$.
The determinant of $\tilde{\Sigma}$ can also be expressed in terms of $\Lambda$:
\begin{align}
& \det \begin{bmatrix}
\Sigma_{AA} & - \Sigma_{AB} \\ \Sigma_{BA} & I - \Sigma_{BB} 
\end{bmatrix} \notag \\
=& \;  \frac{1}{ \det\bigl[ \Lambda_{BB} (\Lambda_{BB} - I)^{-1}  \bigr] \det\bigl[ \Lambda_{AA} - \Lambda_{AB} (\Lambda_{BB} - I)^{-1} \Lambda_{BA} + \Lambda_{AB} (\Lambda_{BB} - I)^{-1} \Lambda_{BB}^{-1} \Lambda_{BA} \bigr] }, \notag \\
=&\;  \frac{1}{ \det \bigl[ \Lambda_{BB} (\Lambda_{BB} - I)^{-1} \bigr] \det \bigl[ \Lambda_{A|B} \bigr] }, \notag \\
=& \; \frac{ \det (\Lambda_{BB} - I) }{ \det \Lambda}.
\end{align}

\section{Properties of Grassmann numbers \label{sec: grassmann_appendix} }
In this appendix, we introduce Grassmann numbers for readers who are not familiar with this algebra.
$p$-dimensional Grassmann numbers $\xi_i, i \in \{ 1, 2, \dots, p\}$ are anticommuting numbers obeying the following anticommutation relation:
\begin{align}
\{ \xi_i, \xi_j\} \equiv \xi_i \xi_j + \xi_j \xi_i = 0.
\end{align}
From this definition, we can immediately read the nilpotency, $\xi_i^2 = 0$.
A monomial of Grassmann numbers can be divided into those consisting of products of an even number of Grassmann numbers (Grassmann even) and those consisting of products of an odd number (Grassmann odd).
Monomials of Grassmann even commute with all the other Grassmann numbers.
A linear transformation of Grassmann numbers by a transformation matrix $C_{ij}$ is also a Grassmann number and satisfied the anticommutation relation,
\begin{align}
\eta_i \equiv \sum_{j=1}^p C_{ij} \xi_j, \\
\{\eta_i, \eta_j\} = 0.
\end{align}

Functions of Grassmann variables are defined by power series expansions.
Because of the nilpotency of Grassmann numbers, the Taylor series expansion terminates at low order.
For example, in the case of two-dimensional Grassmann variables, the Grassmann function is generally expressed in the following form:
\begin{align}
f(\xi) =& c_0 + c_1 \xi_1 + c_2 \xi_2 + c_{3} \xi_1 \xi_2, \notag \\
=& c_0 + c_1 \xi_1 + c_2 \xi_2 - c_{3} \xi_2 \xi_1,
\end{align}
where the coefficient $c_i$ is an ordinary number.
In particular, the exponential function terminates at the first order:
\begin{align}
e^{\xi_i} =& 1 + \xi_i, \\
e^{\xi_i \xi_j} =& 1 + \xi_i \xi_j.
\end{align}
For arbitrary Grassmann functions $f(\xi)$ and $g(\xi)$, the Baker-Cambell-Hausdorff formula hold:
\begin{align}
e^{f(\xi)} e^{g(\xi)} = e^{h(\xi)},
\end{align}
where $h(\xi)$ is a formal series in $f(\xi)$ and $g(\xi)$ and iterated commutators of them,
\begin{align}
h(\xi) =  f(\xi) + g(\xi) + \frac{1}{2} \bigl[f(\xi) , g(\xi) \bigr] + \frac{1}{12} \bigl[f(\xi), \bigl[f(\xi), g(\xi)\bigr]\bigr] - \frac{1}{12} \bigl[ g(\xi), \bigl[f(\xi), g(\xi)\bigr]\bigr] + \cdots,
\end{align}
$[f(\xi), g(\xi)] \equiv f(\xi) g(\xi) - g(\xi) f(\xi)$ is the commutator and $\cdots$ indicates terms involving higher commutators of $f(\xi)$ and $g(\xi)$.
For a product of two Grassmann numbers $\xi_i \xi_j$, the Baker-Cambel-Hausdorff formula reduces to
\begin{align}
e^{\xi_i \xi_j} e^{g(\xi)} = e^{ \xi_i \xi_j + g(\xi)},
\end{align}
since a monomial of Grassmann even commutes with all the other Grassmann numbers.

We can define the differentiation of Grassmann variables in a formal sense.
Naive definition of differentiation conflicts with the anticommutation relation.
We define the differentiation as
\begin{align}
\frac{\partial}{\partial \xi_i} \bigl( 1 \bigr) =& 0, \\
\frac{\partial}{\partial \xi_i} \bigl(\xi_j \bigr)=& \delta_{ij},
\end{align}
with the following anticommuting properties:
\begin{align}
\Bigl\{\xi_i, \frac{\partial}{\partial \xi_j} \Bigr\} =& \delta_{ij}, \\
\Bigl\{\frac{\partial}{\partial \xi_i}, \frac{\partial}{\partial \xi_j} \Bigr\} =& 0.
\end{align}
When the derivative acts on a monomial of Grassmann numbers, it is understood that the variable to be differentiated is brought to the leftmost position:
\begin{align}
\frac{\partial}{\partial \xi_4} \; ( \xi_1 \xi_2 \xi_3 ) \xi_4 =& - \frac{\partial}{\partial \xi_4}   \; \xi_4 ( \xi_1 \xi_2 \xi_3 ) , \notag \\
=& -  \xi_1 \xi_2 \xi_3,
\end{align}
which is also called the left derivative.
The anticommutation relation between $\xi_i$ and $\partial/\partial \xi_j$ is consistent with the left derivative.
The differentiation of the above definition satisfies the following rules, linearity, the graded Leibniz rule and the chain rule, respectively:
\begin{align}
 \frac{\partial}{\partial \xi_i } \Bigl[ a f(\xi) + b g(\xi) \Bigr] =&  a \frac{\partial f}{\partial \xi_i } + b \frac{\partial g}{\partial \xi_i }, \\
\frac{\partial }{\partial \xi_i} \Bigl( B_r B_s \Bigr) = & \left(\frac{\partial B_r}{\partial \xi_i} \right) B_s + (-1)^r B_r \left( \frac{\partial B_s}{\partial \xi_i} \right), \\
\frac{\partial }{\partial \xi_i} f\bigl(g(\xi), h(\xi) \bigr) =&  \frac{\partial g}{\partial \xi_i} \frac{\partial f}{\partial g} + \frac{\partial h}{\partial \xi_i} \frac{\partial f}{\partial h},
\end{align}
where $B_r$ is a monomial consisting of a product of $r$ Grassmann numbers, i.e., a monomial of degree $r$.

We can also define the integration of Grassmann variables.
We regard the properties that hold for definite integrals of ordinary numbers, $\int_{-\infty}^{\infty} dx$, as fundamental properties of integrals.
Hence, we require the following properties, linearity, integration by parts and the shift invariance of integration variables, respectively, for the Grassmann integral:
\begin{align}
\int d\xi_i  \bigl[ a f(\xi) + b g(\xi) \bigr] = & a \int d \xi_i  f(\xi) + b \int d\xi_i g(\xi), \\
 \int d\xi_i \frac{\partial }{\partial \xi_i } \bigl[ f(\xi) \bigr] = & 0, \\
\int d \xi_i f(\xi_i + \xi_j) =&  \int d \xi_i f(\xi_i).
\end{align}
We find that these properties are satisfied if we define the Grassmann integration as the same operation as the Grassmann differentiation:
\begin{align}
\int d\xi_i f(\xi) \equiv \frac{\partial}{\partial \xi_i} f(\xi).
\end{align}
More specifically, the integral is performed as follows:
\begin{align}
 \int d\xi_i \,  ( 1 ) =&  0, \\
 \int d \xi_i \;  \xi_i  =& 1, \\
 \int d \xi_1 d \xi_2 \,  f(\xi) =& - \int d\xi_2 d \xi_1 \, f(\xi),
\end{align}
where the innermost integral is understood to be performed first in the multiple integral.
The above definition of the Grassmann integral corresponds to the following sign convention~\cite{Berezin1966},
\begin{align}
\int d\xi_1 d\xi_2 \cdots d\xi_p \; \xi_p \cdots \xi_2 \xi_1 = + 1.
\end{align}

Here we enumerate useful formulas that hold for Grassmann variables.
For a linear transformation by a nonsingular transformation matrix $A_{ij}$,
\begin{align}
\xi_i =& \sum_{j=1}^p A_{ij} \eta_j,
\end{align}
the differentiation operator transforms as
\begin{align}
\frac{\partial}{\partial \xi_i} =& \sum_{j=1}^p \frac{\partial \eta_j}{\partial \xi_i} \frac{\partial }{\partial \eta_j}, \notag  \\
=& \sum_{j=1}^p A_{ji}^{-1} \frac{\partial}{\partial \eta_j}.
\end{align}
Then, the monomial of Grassmann variables is transformed as
\begin{align}
\xi_{1} \xi_{2} \cdots \xi_{p} =& \sum_{i_1, i_2,\dots i_p=1}^p A_{1 i_1} A_{2 i_2} \cdots A_{p i_p} \;  \eta_{i_1} \eta_{i_2} \cdots \eta_{i_p} , \notag \\
=& \sum_{i_1, i_2, \dots, i_p=1}^p A_{1 i_1} A_{2 i_2} \cdots A_{p i_p} \; \varepsilon_{i_1 i_2 \cdots i_p} \; \eta_{i_1} \eta_{i_2} \cdots \eta_{i_p}, \notag \\
=& \det A \; \eta_1 \eta_2 \cdots \eta_p ,
\end{align}
where $\varepsilon_{i_1 i_2 \cdots i_p} $ is the Levi-Civita symbol defined by
\begin{align}
\varepsilon_{i_1 i_2 \cdots i_p} = \begin{cases} + 1, \hspace{0.2cm} \text{if } (i_1, i_2, \dots, i_p) \text{ is  an even  permutation of  } (1,2, \dots, p) \\ -1, \hspace{0.2cm} \text{if }  (i_1, i_2, \dots, i_p)  \text{ is  an  odd  permutation of } (1,2, \dots, p) \\ 0 , \hspace{0.52cm} \text{otherwise}
\end{cases}.
\end{align}
In a similar way, the multiple integral is transformed as
\begin{align}
\int d\xi_1 d\xi_2 \cdots d\xi_p =& \frac{\partial}{\partial \xi_1} \frac{\partial}{\partial \xi_2} \cdots \frac{\partial}{\partial \xi_p}, \notag \\
=& \frac{1}{\det A} \frac{\partial }{\partial \eta_1} \frac{\partial}{\partial \eta_2} \cdots \frac{\partial }{\partial \eta_p}, \notag \\
=& \frac{1}{\det A} \int d\eta_1 d\eta_2 \cdots d\eta_p.
\end{align}
The Dirac delta function of Grassmann variables is represented by the linear function:
\begin{align}
\delta(\xi_i - \xi_j) = & (\xi_i - \xi_j ), \\
\int d\xi_i \delta(\xi_i - \xi_j )  f(\xi_i) = & f(\xi_j).
\end{align}
The delta function is also represented by the Fourier transform:
\begin{align}
\delta(\xi_i) = \int d\xi_j \exp(\xi_j \xi_i).
\end{align}
For $p$-dimensional Grassmann variables $\bar{\xi}_i$ and $\xi_i$, the Gaussian integral over the Grassmann variables is given by
\begin{align}
\int d\xi_1 d\bar{\xi}_1 d\xi_2 d\bar{\xi}_2 \cdots d\xi_p d\bar{\xi}_p \; \exp \Bigl\{\sum_{i,j=1}^p \bar{\xi}_i A_{ij} \xi_j \Bigr\} = \det A.
\end{align}
This formula can be derived using a linear transformation of Grassmann variables and the integral representation of the delta function.

\bibliographystyle{unsrt}
\bibliography{multivariate_binary}

\end{document}